\documentclass[twocolumn,showpacs,preprintnumbers,amsmath,amssymb,aps,pre]{revtex4}
\usepackage{graphicx}
\usepackage{dcolumn}
\newcommand{\be}{\begin{equation}}
\newcommand{\ee}{\end{equation}}
\newcommand{\bea}{\begin{eqnarray}}
\newcommand{\eea}{\end{eqnarray}}

\newcommand{\zzeta}{{\boldsymbol\zeta}}
\newcommand{\Angstrom}{{\buildrel _{\circ} \over {\mathrm{A}}}}

\newcommand{\mean}[1]
{\mbox{$\langle{#1}\rangle$}}

\begin{document}
\title{Imposing strong correlated energy spread on relativistic bunches with transverse deflecting cavities}


\author{Nikolai Yampolsky}
\email[]{nyampols@lanl.gov}
\affiliation{Los Alamos National Laboratory, Los Alamos, New Mexico, 87545, USA}

\author{Alexander~Malyzhenkov}
\email[]{malyzh@hawaii.edu}
\affiliation{Los Alamos National Laboratory, Los Alamos, New Mexico, 87545, USA}
\affiliation{Northern Illinois Center for Accelerator \& Detector Development and Department of Physics, Northern Illinois University, DeKalb IL, USA}

\author{Evgenya I. Simakov}
\affiliation{Los Alamos National Laboratory, Los Alamos, New Mexico, 87545, USA}


\begin{abstract}
We demonstrate a novel scheme for imposing and removing large energy chirp on relativistic electron bunches relying on the transverse-to-longitudinal mixing accomplished with a set of transverse deflecting cavities. The working principles of the new concept are explained using the first order matrix formalism in the approximation of the linear single-particle dynamics. The scheme demonstrates versatility regarding the average beam energy, and hence, positioning along the accelerator beamline. The performance of the scheme is numerically investigated with the {\sc elegant} particle tracking code. It is shown that the associated nonlinear effects, causing emittance deterioration for the extreme quantities of the total energy spread can be effectively compensated by optimization of the Twiss parameters  while relying on the eigen-emittance analysis. The impact of the longitudinal space charge effects on the beam dynamics in the proposed scheme is also investigated.
\end{abstract}

\maketitle

\section{Introduction}\label{sec:Intro}
Relativistic electron accelerators are used for high energy physics studies \cite{KEKB,CEBAF,ILC} and as light sources~\cite{APS, NSLS, FLASH, FERMI, LCLS, Spring-8, XFEL,PAL-XFEL,SWISSFEL}. These applications
require high quality and large peak current electron bunches. Such beams are subject to coherent synchrotron radiation (CSR)~\cite{CSR-JETP,CSR,Saldin_CSR,Bruce_Tor} which results in the variation of the beam energy
along the bunch and degradation of the beam quality. As a result, the applications demanding the highest beam quality
rely on the linear accelerators~\cite{LCLS,Spring-8,XFEL,SWISSFEL} which have smaller number of bending magnets compared to circular accelerators~\cite{ILC}.

The design of these machine results in the need of several bunch compressors (BC) at different energies for achieving bunches with large peak current. Compression is achieved through imposing
energy slew along the bunch and passing it through the dispersive beam element such as chicane consisting of four magnetic bends~\cite{Saldin_CSR}.
Electrons in the bunch have energy chirp. The energy of electrons is larger at the tail of the bunch and smaller at the head of the bunch.
Electrons with higher energy travel closer to the axis in magnetic bends due to higher
relativistic $\gamma$ factor and their time of travel through the chicane is smaller compared to that of the low energy electrons. As a result, higher energy electrons at the tail of the bunch catch up with electrons at the head and the chirped bunch compresses in time.

Conventionally, the bunches are chirped while being accelerated off-crest in the RF field.
For the off-crest acceleration, longitudinal variation of the RF fields results in difference in accelerating
gradients along the bunch. However, the accelerating gradient for off-crest acceleration is lower than on-crest.
Imposing the correlated energy spread on the beam in the off-crest acceleration results in the trade-off between the accelerating gradient ($\sim\cos\phi_0$) and the induced energy chirp which is defined by the $R_{65}\sim\sin\phi_0$ element of the linear transform matrix $R$:
\begin{equation}
R_{65}=-\frac{eE_0 k \sin\phi_0}{\gamma mc^2}s,
\end{equation} 
where $E_0$ and $k$ are respectively the amplitude and wave vector of the RF field, $\phi_0$ is the RF phase, $\gamma$ is the normalized average energy of the bunch,
$s$ is the effective length of the RF structure, $m$ and $e$ are the mass and charge of the electron, and $c$ is the speed of light. The larger energy chirp in the beam requires off-crest acceleration at lower gradient and results in the loss of the final beam energy in a given
linac section. For example, the off-crest acceleration results in
the loss of 1.575 GeV of the beam energy at Linac Coherent Light Source (LCLS)~\cite{LCLS_design} ($-43^{\circ}$ RF phase between the first bunch compressor (BC1) at 250 MeV and
the second bunch compressor (BC2) at 4.54 GeV)
and 1.5 GeV of the beam energy at Matter-Radiation Interactions in Extremes (MaRIE) linac~\cite{MARIE_update_2015} (-20$^{\circ}$ RF phase between BC1 at 250~MeV and BC2 at 1~GeV). The off-crest acceleration is particularly inefficient for chirping short bunches since the difference between the accelerating fields at the head and the tail of the
bunch is proportional to the bunch length, $\sigma_{\Delta\gamma/\gamma}=R_{65}\sigma_z$.
As a result, additional linac's length is required in off-crest acceleration to compensate for the drop in the accelerating gradient. 

In this paper we propose an alternative method of chirping relativistic electron bunches employing transverse-to-longitudinal mixing accomplished by a set of transverse deflecting cavities (TCAVs).
Such cavities operate in the TM$_{110}$ deflecting mode. Both electric and magnetic fields are zero on axis at zero RF phase and the motion of the reference particle characterizing the average beam dynamics is not affected by TCAVs.
The magnetic field in a TCAV varies longitudinally providing angular deflection to electrons which depends on their longitudinal position in the bunch.
The transverse variation of the longitudinal electric field provides different accelerating gradients at different transverse locations within
the bunch and may be utilized for imposing correlated energy variations within the bunch. The efficient longitudinal chirping of the bunch would require strong correlation between the longitudinal and transverse positions of particles in the bunch. In
this case the transverse variation of the electrons' energy would also manifest as a longitudinal chirp.
The required $x-z$ correlation inside the bunch can be created using another TCAV,
which deflects electrons at some angle proportional to their longitudinal coordinate, followed by the vacuum drift space which translates angular divergence into the
transverse displacement. Similar methods of imposing $x-z$ correlation are routinely used for longitudinal diagnostics of relativistic beams~\cite{TCAV-diag}.

The proposed  {\bf T}ransverse deflecting {\bf C}avity {\bf B}ased {\bf C}hirper (TCBC) exploits a great flexibility in manipulations with the
transverse bunch size unlike its length, which is fixed in relativistic beams under acceleration. As a result, the difference in accelerating fields between two ends
of the bunch is determined by its transverse size which can easily be made much larger than its length.

\section{Linear analysis}
\label{sec:Linear_A}
In this section we introduce a TCBC beamline consisting of three TCAVs. We analyze this scheme in the approximation of the linear single-particle dynamics and demonstrate that it is capable of imposing large longitudinal energy chirp on relativistic bunches.

\begin{figure}[ht]
\center
\includegraphics[width=3.375in,keepaspectratio]{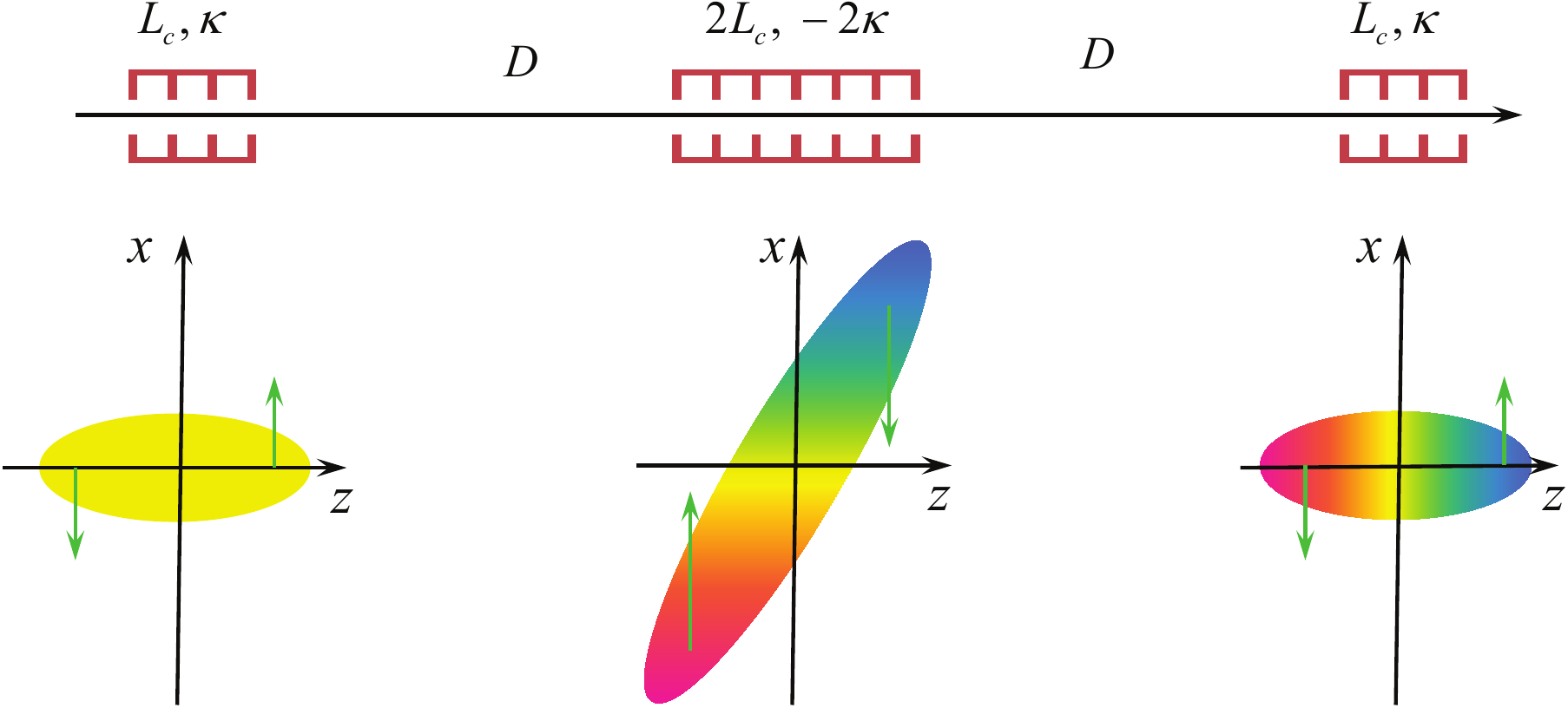}
\caption{[Color online] The schematics of the TCBC beamline (above) consisting of three TCAVs for imposing longitudinal energy chirp in relativistic bunches. Subplots bellow show the
$x-z$ electron bunch distribution inside each TCAV. Green arrows in these plots show $z$-dependent deflection provided by each cavity and color shows the particle energy at the corresponding location (larger energy is shown as red and lower energy as blue).}
\label{scheme}
\end{figure}

The schematics of the beamline is shown in Fig.~{\ref{scheme}} and consists of three cavities, namely TCAV1, TCAV2, and TCAV3, separated by equal vacuum drifts of length $D$. The side cavities are identical, their lengths are $L_c$ and the normalized deflecting potential (also known as the ``cavity strength") is $\kappa=eV_\perp/(mc^2\gamma a)$, where $a$ is the characteristic cavity size on the order of the RF wavelength and $V_\perp$ is the TCAV deflecting voltage. The middle cavity is twice as long as the side cavities, and therefore, has twice the deflecting voltage compared to the side cavities assuming the same field amplitude in all TCAVs~\cite{note1}. Moreover, the middle cavity has RF phase shift of $\pi$
compared to the side cavities at the time of the bunch arrival which manifests as the deflecting voltage of an opposite sign.

Beam dynamics of the bunch along the TCBC beamline can be explained as follows. The bunch acquires $z$-dependent angular deflection in the first cavity and starts expanding in the drift space after the cavity. That expansion results in a strong $x-z$ correlation inside the bunch and its transverse size becomes much larger than its length. The middle cavity changes
electrons' energy based on their transverse position which manifests as longitudinal energy chirp in a bunch with strong $x-z$ correlation. At the same time, the middle cavity produces 
the $z$-dependent angular kick to electrons which is twice as strong as that of the TCAV1 and has the opposite sign. As a result, the bunch is focused back in the downstream
drift and restores its original shape when it arrives to the last cavity. TCAV3 predominantly compensates the residual $z$-dependent angular deflection in the beam.
The beamline introduces strong correlations between the longitudinal and transverse phase spaces which are used to impose the longitudinal energy chirp on the beam relying on the
transversely-dependent accelerating field. All introduced correlations are removed at the end of the beamline and the entire impact of the TCBC is solely the longitudinal
energy chirp, mainly imposed in the middle cavity. 
 
We analyze the performance of the proposed TCBC scheme within the limits of linear beam optics formalism under the assumption of the quasi-mono-energetic bunch with a small angular divergence, and the geometric size much smaller than the RF wavelength. The beam is represented as an ensemble of electrons characterized by coordinates in the phase space. Each
element of the beamline affects the electron phase space through the beam transform matrix $M$. Our
analysis is limited to the 4-dimensional (4D) phase space $(x, x^\prime, z, \Delta \gamma/\gamma)$ since the dynamics in the
transverse phase space $(y, y^\prime)$ is decoupled from other dimensions and manifests as a simple drift. This representation is simpler compared to the full 6D 
description using the full beam transport matrix R. The transform matrices of the drift space $M_d$ and the TCAV $M_c$ \cite{EEX} are:
\bea
M_{d}(D)&=&\left(
\begin{array}{cccc}
1&D&0&0\\
0&1&0&0\\
0&0&1&0\\
0&0&0&1
\end{array}\right),\\
M_{c}(\kappa,L_c)&=&\left(
\begin{array}{cccc}
1&L_c&\frac{\kappa L_c}{2}&0\\
0&1&\kappa&0\\
0&0&1&0\\
\kappa&\frac{\kappa L_c}{2}&\frac{\kappa^2 L_c}{6}&1
\end{array}\right).
\eea
The overall transform matrix of the entire beamline can be found as the multiplication product of its structural components
$M_{TCBC}=M_c(\kappa,L_c) M_d(D) M_c(-2\kappa,2L_c) M_d(D) M_c(\kappa,L_c)$ that results in:
\be
\label{TCBC}
M_{TCBC}=\left(
\begin{array}{cccc}
1&2D+4L_c&0&0\\
0&1&0&0\\
0&0&1&0\\
0&0&-\frac{2}{3}\kappa^2(3D+2L_c)&1
\end{array}\right).
\ee
The transverse and the longitudinal phase spaces are decoupled at the end of the beamline and the start-to-end transform matrix is block-diagonal. The dynamics in the
transverse phase space represents the drift corresponding to the overall beamline's length and the $M_{43}\equiv R_{65}<0$ matrix element results in the compressing chirp in the longitudinal phase space since the head of the bunch ($z>0$) looses energy ($\Delta\gamma<0$) while the back of the bunch
($z<0$) gains energy ($\Delta\gamma>0$). 

Equation~(\ref{TCBC}) shows clear advantages of the proposed TCBC scheme over conventional off-crest acceleration. The $R_{65}$ element scales
quadratically with the cavity's strength $\kappa$ as opposed to the case of the off-crest acceleration which scales linearly with $\kappa$, yet the sign of the chirp in this novel scheme cannot be simply changed by delaying the RF phase by 180$^\circ$ (this phase delay in the dipole mode is equivalent to the rotation of the cavity by 180$^\circ$ in the ($x,\;y$) plane, see Sec.~\ref{sec:Dechirper}). The quadratic scaling indicates that the TCBC is a cost efficient solution for
imposing large energy chirps. Furthermore, the imposed chirp in the TCBC scheme scales linearly with the drift lengths between TCAVs. This allows to reduce the cost of the system by
replacing expensive RF cavity sections with cheap vacuum drifts.

In the limit of a large vacuum drift compared to the length of TCAVs, the chirp imposed by the TCBC can be approximated
as $R_{65}\approx-2\kappa^2D$. The correlated energy variation in the bunch is acquired predominantly in the middle deflecting cavity in which the transverse beam size is the largest due to strong transverse-to-longitudinal correlations. The transverse beam size in TCAV2 is on the order of $\sigma_{x,TCAV2}\approx(\kappa D)\sigma_z$, and the chirp imposed by the TCBC beamline is on the order of
$R_{65}\approx-2\kappa(\sigma_{x,TCAV2}/\sigma_z)$. Consequently, the efficiency of the TCBC over the conventional off-crest acceleration in the cavity of the same strength $-2\kappa$ (defined as normalized voltage, similar to that of the deflecting cavity)
is increased by a factor of $(\sigma_{x,TCAV2}/\sigma_z)$. That explains the underlying idea of imposing chirp using transversely large rather than longitudinally short
bunches. The transverse beam size is physically limited by the beam pipe radius and by the RF wavelength, and can be extended into mm- and cm- range while the bunch length is required to be short and it is typically on the 
order of 0.1 mm in modern X-ray free-electron lasers (FELs). Summarizing the above, the TCBC scheme can be 10-100 times more efficient compared to conventional off-crest acceleration.

\section{Design of TCBC for MaRIE FEL}\label{sec:Chirper}

The MaRIE x-ray FEL proposed at Los Alamos National Laboratory (LANL) is currently planned to lase at the wavelength of $0.3{\Angstrom}$ \cite{MaRIE}.
The electron beam will be accelerated in a 1.3 GHz superconducting linac reaching the energy of 12 GeV. The linac includes two
bunch compression stages at 250 MeV and 1 GeV (referred further as BC1 and BC2). High requirements on the beam quality require imposing an excessive energy chirp of $\pm$5~MeV on the bunch in order to reduce the effect of CSR during
compression in BC2 and preserve the required normalized emittance of $\varepsilon_n=0.1\;\mu$m.
Low RF frequency and close energies of the compression stages result in a large off-crest phase of RF which results in the accelerating gradient
more than a factor of 2 smaller than the peak value. This approach is not cost efficient. Moreover, low accelerating gradient results in the increased slice energy growth caused
by the microbunching instability driven by the longitudinal space charge~\cite{Saldin_Klystron,Huang_Microbunch_Inst}.

\begin{figure}[ht]
\center
\includegraphics[width=3.4in]{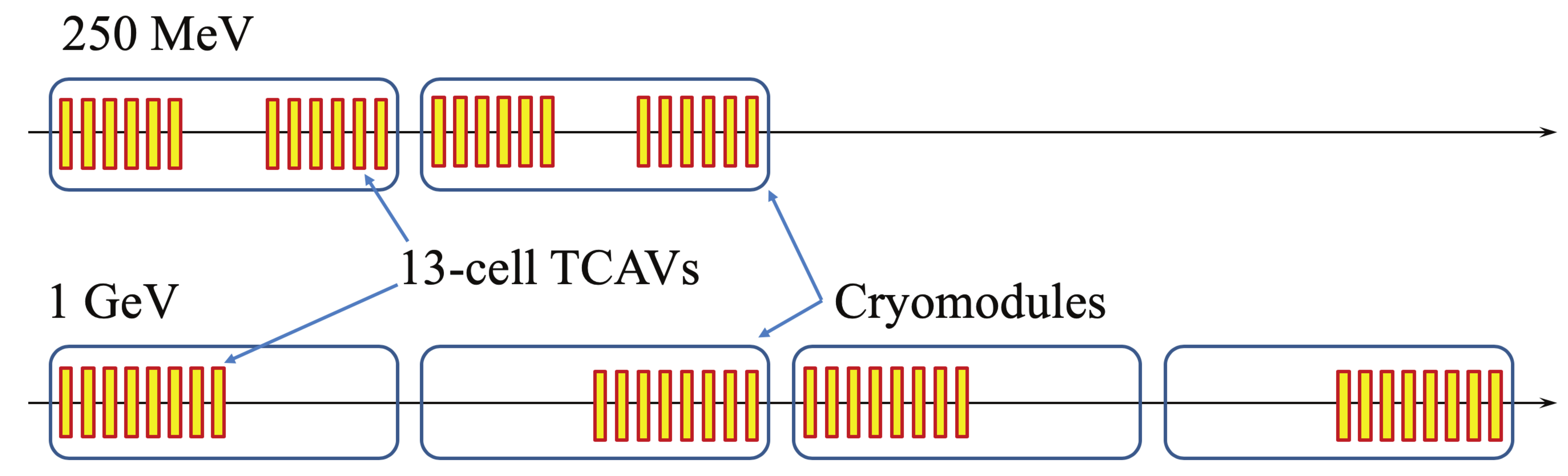}
\caption{[Color online] The layout of TCAVs inside the cryomodules for MaRIE linac. Two design options for TCBC at 250 MeV (above) and 1 GeV (bellow) are shown.}
\label{layout}
\end{figure}

As an alternative solution, we propose to use the TCBC scheme to provide the required energy chirp. The proposed parameters of the electron beam and the TCBC components are listed in Table~\ref{tbl:chirper}. The design was developed through the compromise between the cost efficiency and simplicity of operation. The minimal TCBC cost is achieved when the ratio of drift's and TCAV's
lengths is roughly the same as the ratio between costs of RF infrastructure (including the cost of structures, cryomodules, tunnel, etc.) and vacuum drifts. The simplicity requirement results in
the scheme which can fit within the length of the integer number of standard cryomodules which simplifies replacement. Table~\ref{tbl:chirper} summarizes components of two different designs of the TCBC at the beam energies of
250 MeV (right after BC1) and 1 GeV (right before BC2). Both designs are based on the cryomodules~\cite{ILC-cryo} and 13-cell 3.9~GHz transverse deflecting cavities~\cite{ILC-TCAV} designed for the International Linear Collider (ILC)~\cite{ILC}. The schematic layouts of TCAVs inside the
cryomodules are shown in Fig.~\ref{layout}.
\begin{table}[ht]
\begin{ruledtabular}
\caption{TCBC parameters for MaRIE X-ray FEL}
\begin{tabular}{lll}
& BC1 & BC2 \\ 
{\bf Electron beam}& & \\ 
Beam energy $E=\gamma mc^2$, MeV& 250 & 1000\\
Bunch charge $q$, pC& 100 & 100\\
Bunch length $\sigma_z$, $\mu$m& 90 & 90\\
Normalized transverse emittances $\epsilon_{nx},~\epsilon_{ny}$, $\mu$m& 0.1 & 0.1\\
Normalized longitudinal emittance $\epsilon_{nz}$, $\mu$m& 5.72 & 5.72\\
Uncorrelated energy spread $\gamma mc^2\sigma_{\Delta\gamma/\gamma,u}$, keV & 32.5 & 32.5\\
Correlated energy spread $\gamma mc^2 R_{65}\sigma_{z}$, MeV&6.7 & 8.1\\
{\bf Beamline components}& & \\
Total number of cryomodules & 2 & 4\\
Total number of 13-cell cavities & 24 & 32\\
Cavity frequency, GHz &3.9 & 3.9\\
Effective RF length in each cavity, m &0.5 &0.5\\
Deflecting voltage in each cavity, MeV&2.5 & 2.5\\
Number of 13-cell cavities per TCAV& 6 & 8\\
Drift length $D$, m& 3.39& 6.58\\
TCAV length $L_c$, m&4.28 & 5.72\\
TCAV strength $\kappa$, $\rm m^{-1}$&0.82 & 0.205\\
$R_{65}$, $\rm m^{-1}$& 298& 90\\
Beamline length, m&25.3&50.6\\
\end{tabular}~\label{tbl:chirper}
\end{ruledtabular}
\end{table}

\subsection{Chirper at 250~MeV.}
\label{sec:250MeV}
\begin{figure}[!hb]
	\center
	\begin{tabular}{rrrr}
	&\includegraphics[height=1.75in]{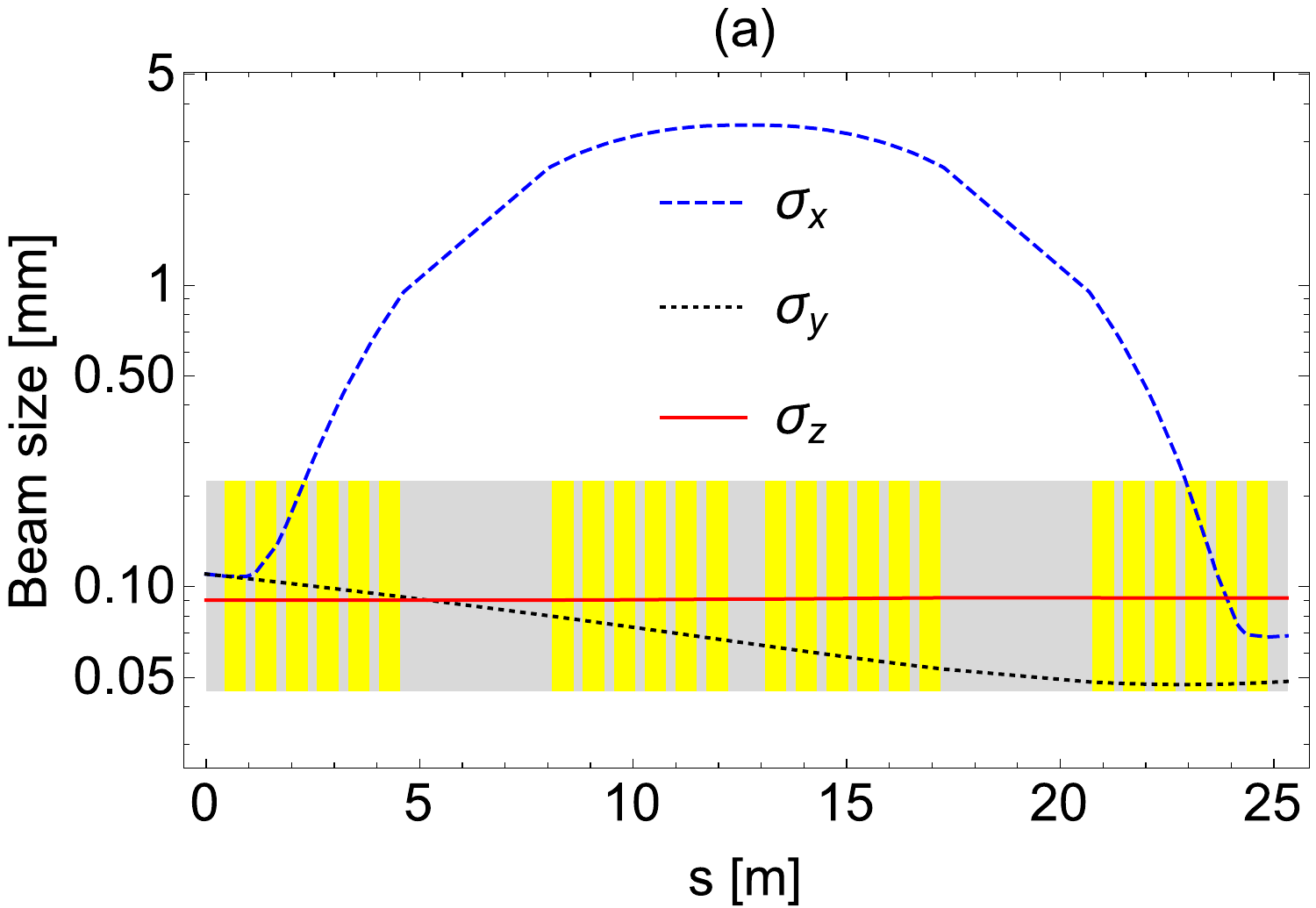}\\
	&\includegraphics[height=1.75in]{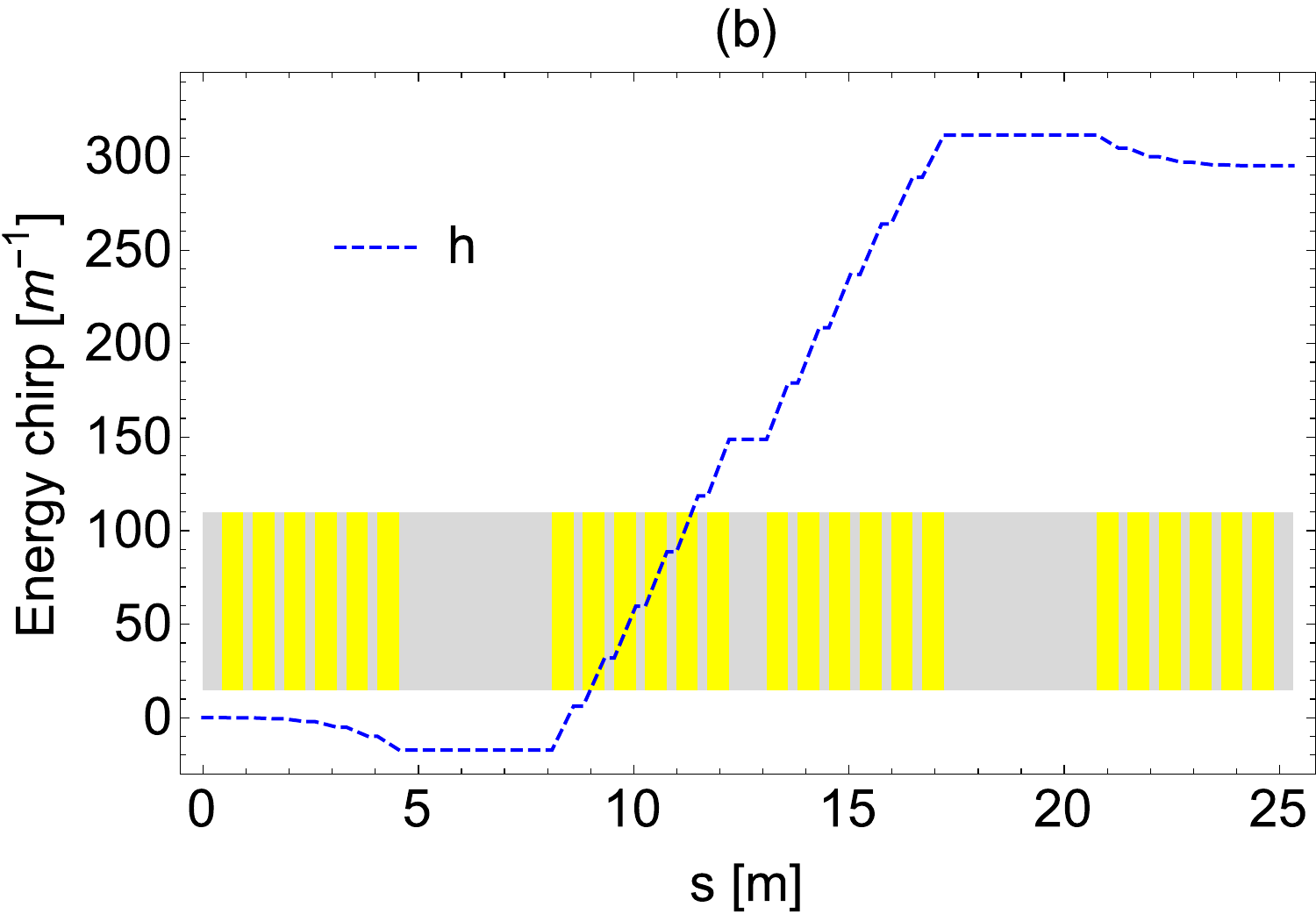}\\
	&\includegraphics[height=1.75in]{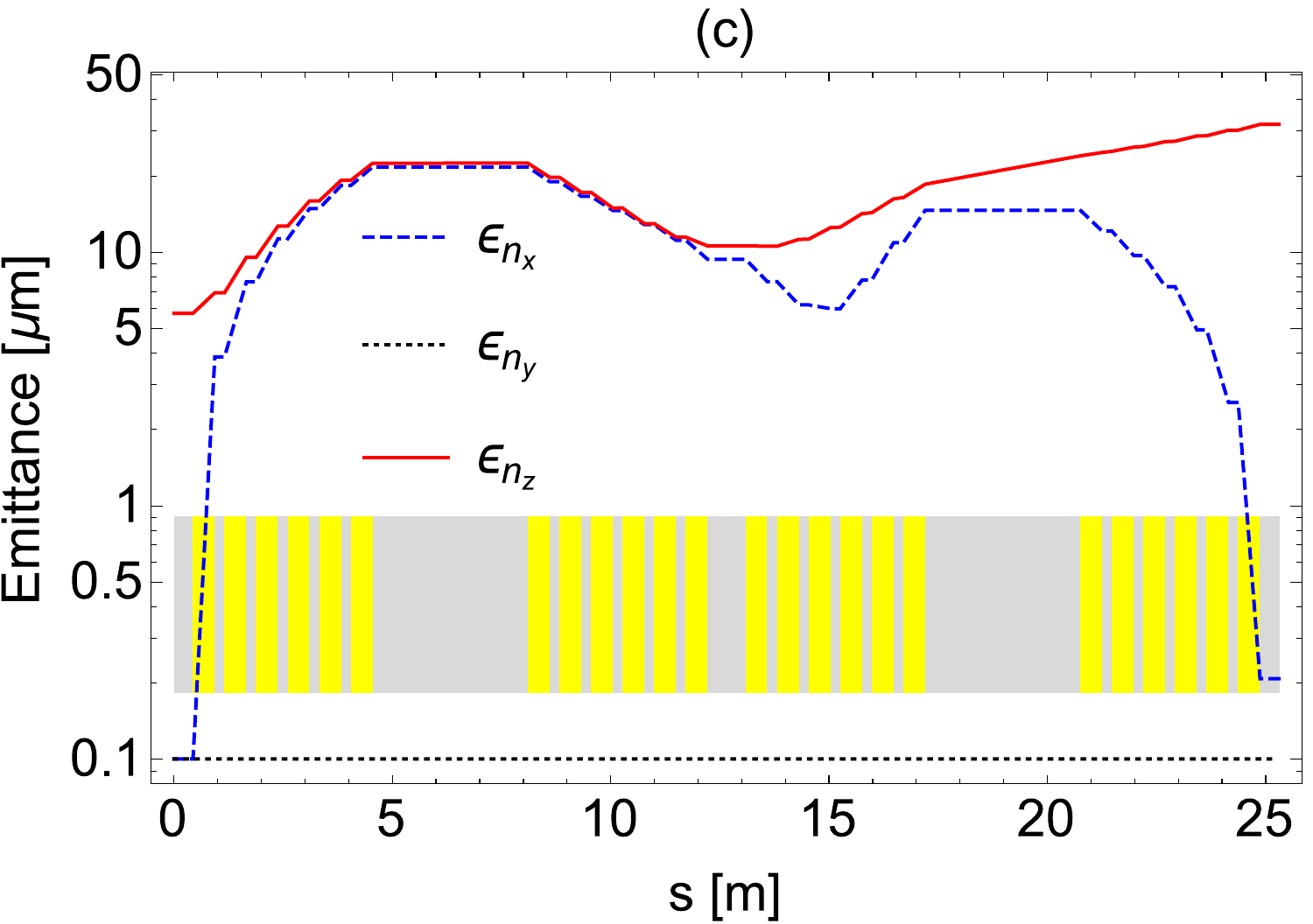}\\
	&\includegraphics[height=1.75in]{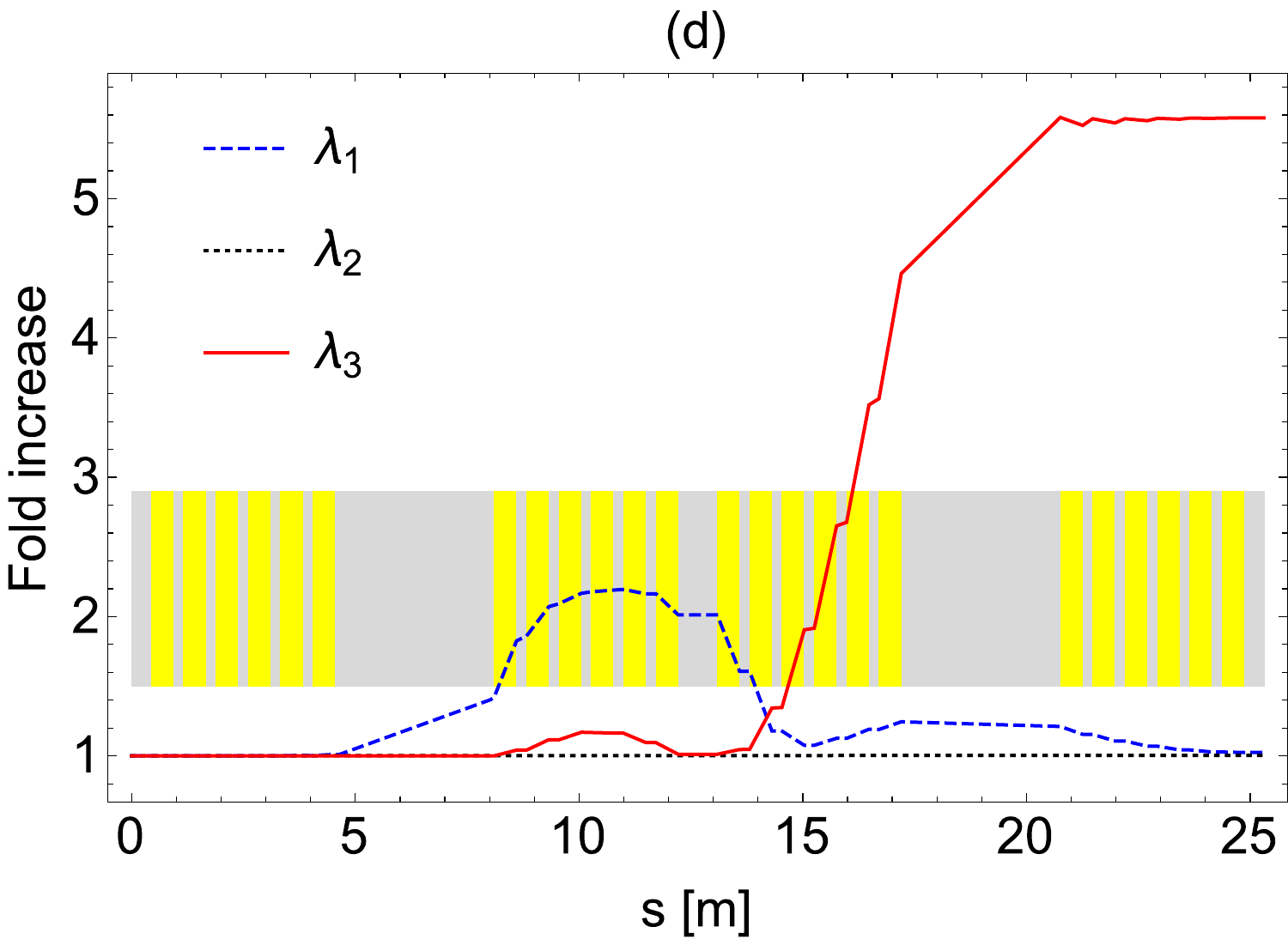}
	\end{tabular}
	\caption{[Color online] Rms beam sizes (a), energy chirp (b), emittances (c), and increase in eigen-emittances (d) along the beamline for the unoptimized Twiss parameters $\beta_x=\beta_y=59$~m and $\alpha_x=\alpha_y=2.1$. Each 13-cell cavity is marked as a yellow rectangle representing its actual geometric length and position along the beamline. }
	\label{fig:emittances_beta_a0}
\end{figure}

The required correlated energy spread of 5~MeV for the 250~MeV beam energy results in $\Delta\gamma/\gamma=0.02$ for particles in the head and the tail of the bunch that is well within the limits of the linear formalism (requires $\Delta\gamma/\gamma\ll1$) presented in Sec.~\ref{sec:Linear_A}. Accordingly, 
the TCBC beamline was simulated using the particle tracking code {\sc elegant}~\cite{Elegant} including nonlinearities of each element up to the 3rd order, and additionally accounting for the longitudinal space charge (LSC) for 100~pC bunch. The later did not result in any visible changes in the beam dynamics in the schemes placed before BC2 which are discussed in this section~\cite{noteLSC}. Twiss parameters are optimized ($\beta_{x,y}=59$~m, $\alpha_{x,y}=2.1$) to result in the smallest increase in the beam's emittance.
Figure~\ref{fig:emittances_beta_a0}(a) demonstrates the evolution of the rms beam sizes along the beamline. In particular, $\sigma_x$ significantly grows in the middle TCAVs due to introduced $z$-dependent angular spread. The transverse-to-longitudinal coupling is removed at the end of the beamline resulting in $\sigma_x$ and $\sigma_y$ that are close to each other and slightly smaller than their initial values due to focusing in a vacuum drift for the chosen input Twiss parameters. The longitudinal beam size is almost unaffected by the transverse-to-longitudinal correlations and remains invariant along the beamline.
The energy chirp is defined using the conventional notations:
\begin{equation}\label{chirp}
h=\frac{\mean{z\cdot\Delta\gamma/\gamma}}{\mean{z^2}}[m^{-1}],
\end{equation}
where $\mean{...}$ denotes the ensemble average.
It is predominantly acquired in TCAV2 (as shown in Fig.~\ref{fig:emittances_beta_a0}(b)) as it was predicted by the linear formalism (see Sec.~\ref{sec:Linear_A}).
Figure~\ref{fig:emittances_beta_a0}(c) demonstrates evolution of the transverse ($\epsilon_{nx}$ and $\epsilon_{ny}$) and longitudinal ($\epsilon_{nz}$) normalized emittances along the beamline. The transverse and longitudinal emittances grow significantly during beam propagation since the motions in the transverse ($x,\;x'$) and longitudinal ($z,\;z'$) phase spaces are highly coupled. The transverse emittance $\epsilon_{nx}$ is 0.21~$\mu$m at the end of the beamline which is roughly a factor of 2 larger than its initial value of 0.1~$\mu$m.

The emittance degradation due to nonlinear and collective effects in the schemes with transverse-to-longitudinal mixing can be investigated by using the eigen-emittance analysis. Linear 6D dynamics of relativistic beams allows 3 invariants characterizing the electron bunch. They can be chosen to be the eigen-emittances $\widetilde{\lambda}_j$, introduced by Dragt~\cite{Dragt, Dragt1}:
\begin{equation}
\det\left(J\Sigma-i\widetilde{\lambda}_jI\right)=0,
\end{equation} 
where $\Sigma=\mean{\zzeta \zzeta^T}$ is the 6D rms beam matrix, $\zzeta=(x,\;x^\prime,\;y,\;y^\prime,\;z,\;\Delta\gamma/\gamma)$ is the 6D coordinate of an electron in the phase space, $J$ is the unit block-diagonal antisymmetric symplectic matrix satisfying $J^2=-I$, and $I$ is the identity matrix. Respectively, the normalized eigen-emittances are defined as: $\lambda_j=\gamma\beta\widetilde{\lambda}_j$.

\begin{figure}[ht]
	\center
	\includegraphics[width=2.5in]{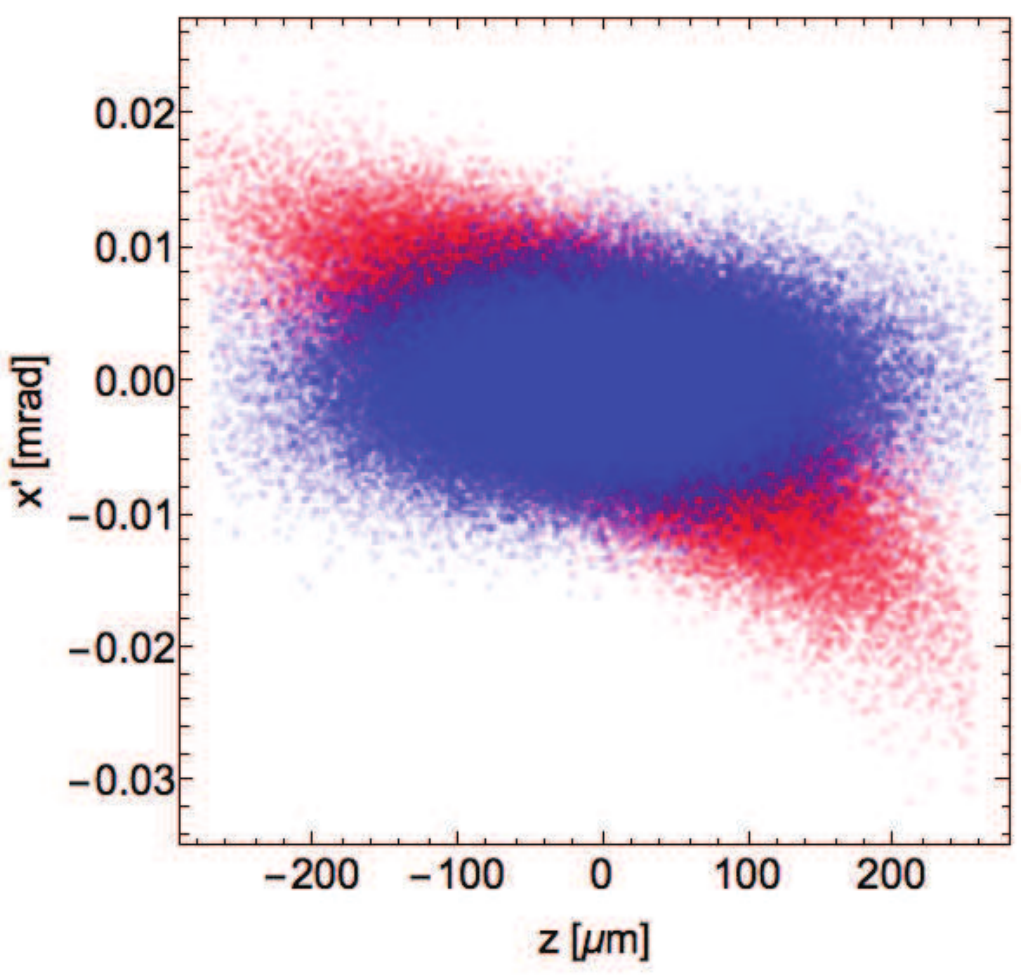}
	\caption{[Color online] Phase spaces at the exit of TCBC at 250 MeV beam energy in the linear (blue) and nonlinear regime (red) for the optimal Twiss parameters $\beta_{x,y}=59$~m and $\alpha_{x,y}=2.1$.}
	\label{Fig:250-PS-optTwiss}
\end{figure}

The evolution of the eigen-emittances along the beamline is shown in Fig.~\ref{fig:emittances_beta_a0}(d).
The transverse dynamics in ($y,\;y'$) phase space is decoupled from other dynamics and results in no visible changes in emittance $\epsilon_{ny}$ and the corresponding eigen-emittance $\lambda_2$ relative to the initial value of 0.1~$\mu$m. 
The longitudinal emittance $\epsilon_{nz}=$31.8~$\mu$m as well as its associated eigen-emittance $\lambda_3=$31.9~$\mu$m at the end of the beamline are significantly larger than their initial values of 5.72~$\mu$m. Longitudinal emittances and eigen-emittances grow simultaneously, which indicates that their increase is caused by nonlinearities of the beamline elements. It is mostly caused by a large imposed energy spread, $\Delta\gamma/\gamma\sim0.2$. 
The eigen-emittance $\lambda_1$ related to the transverse phase space ($x,\;x'$) at the end of the beamline predominantly grows in the middle deflecting cavity and comes back close to its original value in the last cavity resulting in $5\%$ start-to-end growth. 
The increase in the eigen-emittances in the middle of the TCBC beamline and its strong compensation at the end of the beamline is not a real effect. It is an artifact of the {\sc elegant} code which does not use canonical variables to describe the phase space of the beam. At the same time, the significant difference between the transverse emittance $\epsilon_{n_{x}}$ and its corresponding eigen-emittance $\lambda_1$ indicates that the phase space of the beam remains partially coupled between ($x,\;x'$) and ($z,\;\Delta\gamma/\gamma$)
phase planes at the end of the TCBC beamline as shown in Fig.~\ref{Fig:250-PS-optTwiss}. This residual coupling can be explained as follows.
Nonlinear effects slightly change the average beam energy along the beamline corresponding to $\gamma_1=489.225$, $\gamma_2=489.221$, $\gamma_3=489.268$, and $\gamma_4=489.327$ at the entrances to TCAV1, TCAV2, TCAV3, and TCAV4, respectively. As a result, the strengths of each cavity are different from the designed values. As a result, the effective linear matrix of TCBC beamline has uncompensated nonzero off-diagonal elements resulting in the linear transverse-to-longitudinal correlations. 

\begin{figure}[ht]
	\center
		\begin{tabular}{rr}
&	\includegraphics[height=1.75in]{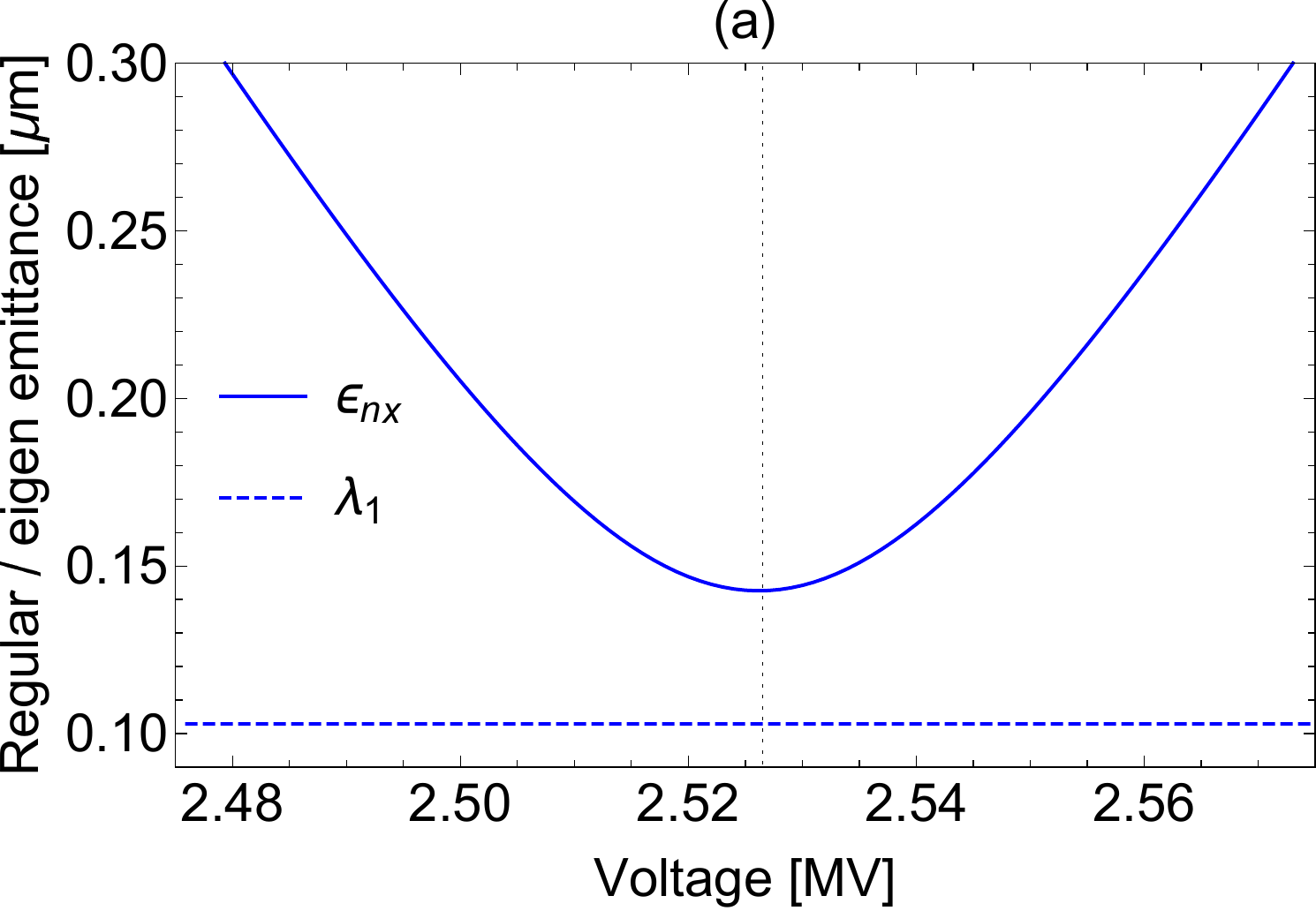}\\
&	\includegraphics[height=1.75in]{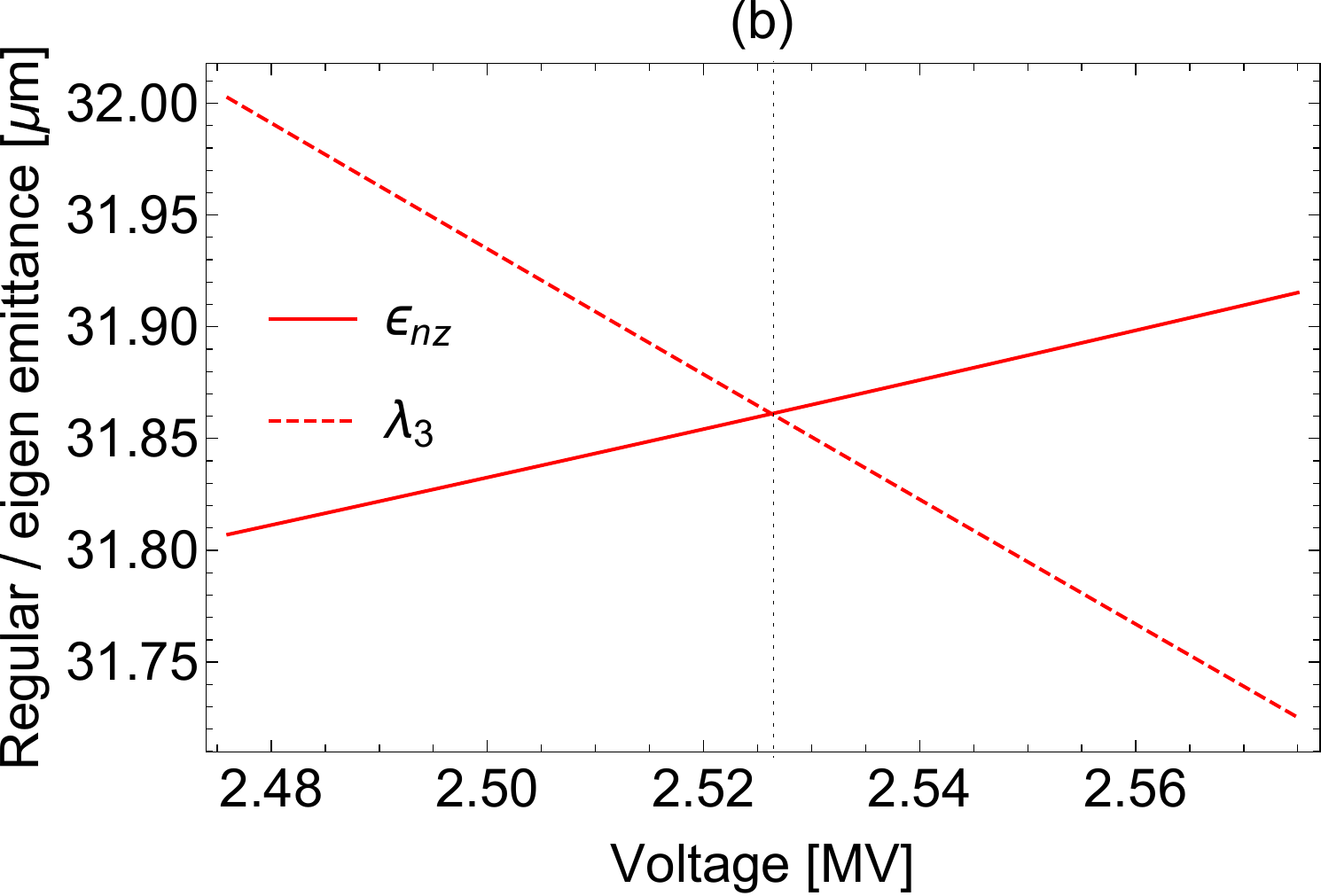}
		\end{tabular}
	\caption{[Color online] (a) Transverse and (b) longitudinal emittances and corresponding eigen-emittances as a function of the voltage in TCAV4. The dotted vertical line shows the optimal voltage at the minimum of the transverse emittance, which occurs when the longitudinal regular and eigen-emittances match.}
	\label{fig:emittances_Last_TDC}
\end{figure}

\begin{figure}[ht]
	\center
	\includegraphics[width=2.5in]{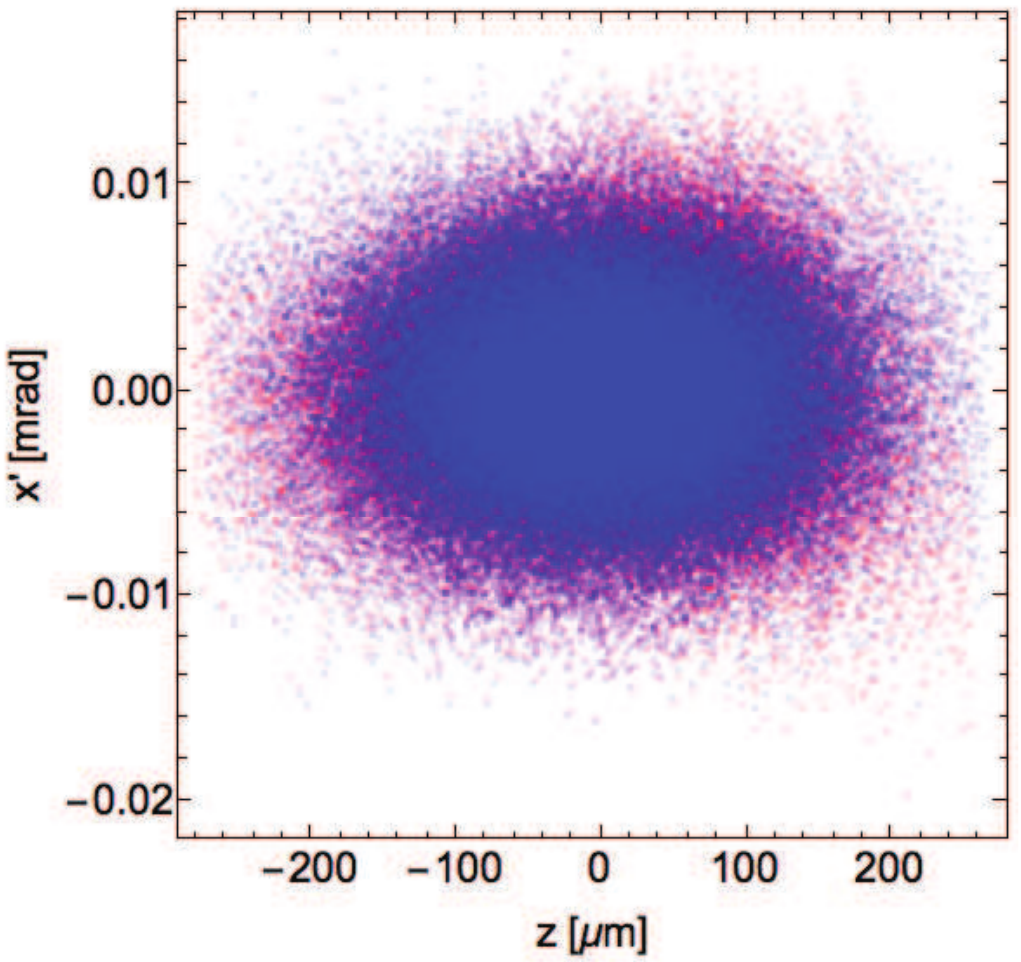}
	\caption{[Color online] Phase spaces at the exit of the TCBC at 250~MeV with the tuned voltage $\Delta V$=+26.5~kV in the last TCAV for the linear (blue) and nonlinear (red) regimes for $\beta_{x,y}=59$~m and $\alpha_{x,y}=2.1$.}
	\label{fig:phase_space_250MeV_nosc_LastTDC}
\end{figure}

The remaining coupling between the longitudinal and transverse phase spaces can be removed through minor adjustment of the TCAV4's voltage from 2.5~MV to 2.5265~MV ($\Delta V$=+26.5~kV), while all other parameters of the beamline are kept unchanged. 
This adjustment does not significantly affect the final chirp (less then 0.12\% reduction). Figure~\ref{fig:emittances_Last_TDC} illustrates that the minimum of the transverse emittance $\epsilon_{nx}$ is achieved at the same voltage when the longitudinal emittance matches the corresponding eigen-emittances. That indicates that the longitudinal and
transverse phase spaces are decoupled in this case, which is illustrated in Fig.~\ref{fig:phase_space_250MeV_nosc_LastTDC}.

The adjustment to the TCAV4 voltage depends nonlinearly on its design value since it should compensate for the nonlinear effects. The TCBC scheme does not depend on
the sign of the deflecting voltage in TCAVs since it is equivalent to rotating the entire beamline by 180 degrees in x-y plane along the z-axis. Therefore,
$sign(\Delta V)=sign(V)$, which indicates that $\Delta V\propto V^3$. Optimization of the TCAV beamline described above resulted in the following empirical scaling for the
adjusted voltage in the TCAV4:
\begin{equation}\label{eq:scaling}
 \Delta V [kV]\approx 1.696\cdot V^3 [MV]. 
 \end{equation}
The empirically found dependence matches the theoretically expected scaling. The exact numerical factor depends on the details of the TCBC beamline, such as: beam energy, TCAV's length, and vacuum drift length.

Twiss parameters can be optimized to minimize the emittance growth due to the quadratic nonlinearities of the beam elements. The optimal set of $\beta_x=150$~m and $\alpha_x=-1.3$ are significantly different from the optimized Twiss parameters prior to correction of TCAV4 voltage ($\beta_{x,y}=59$~m and $\alpha_{x,y}=2.1$), while vertical Twiss parameters are unchanged. The transverse emittance
$\epsilon_{nx}$ and the corresponding eigen-emittance $\lambda_1$ increase by 5\% for the maximum applied chirp of 300 m$^{-1}$. Identical increase in these two parameters indicates that the longitudinal and transverse phase spaces are decoupled by the end of the beamline.

\begin{figure}[ht]
	\center
	\includegraphics[height=0.22\textwidth]{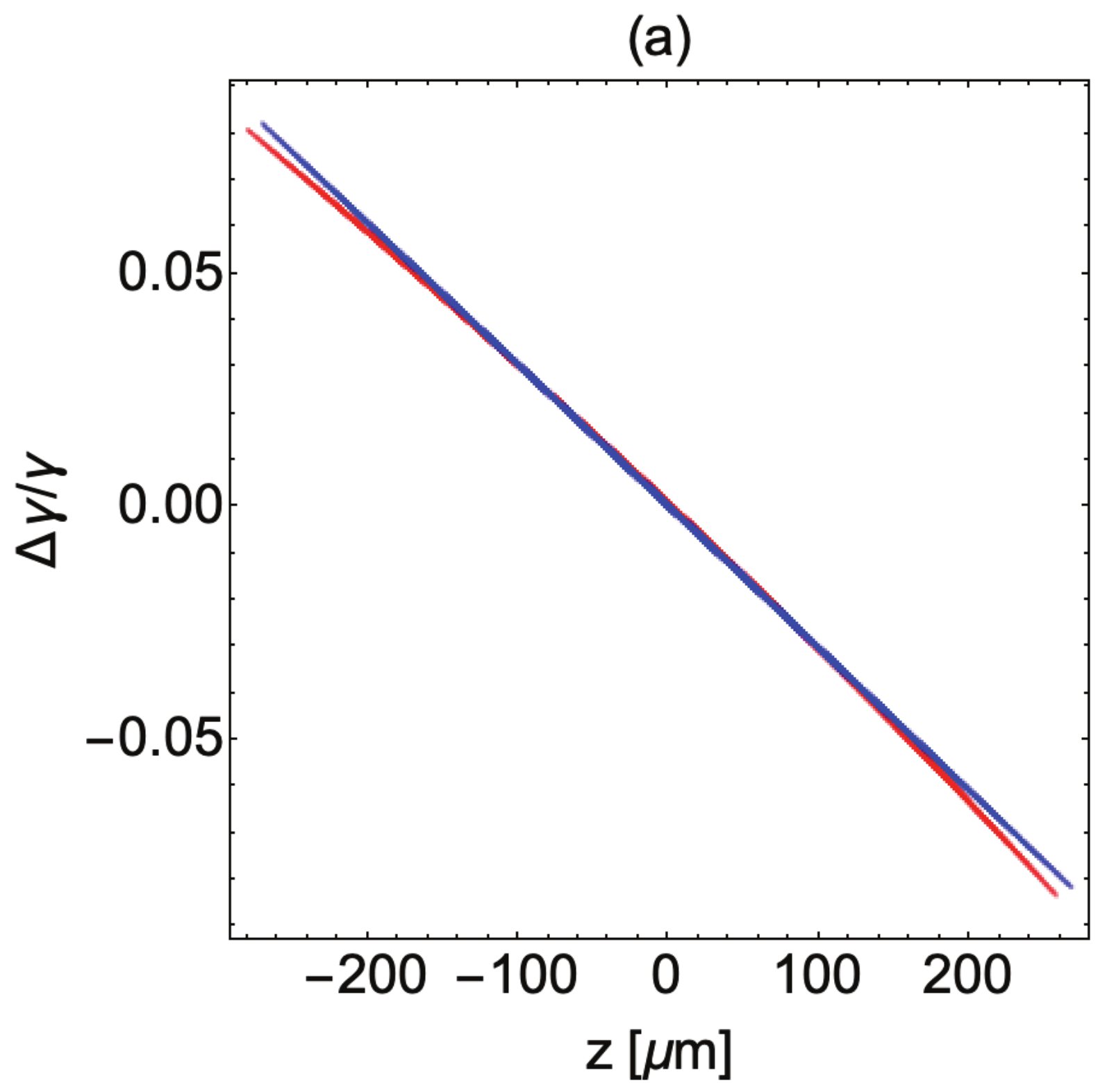}
	\includegraphics[height=0.23\textwidth]{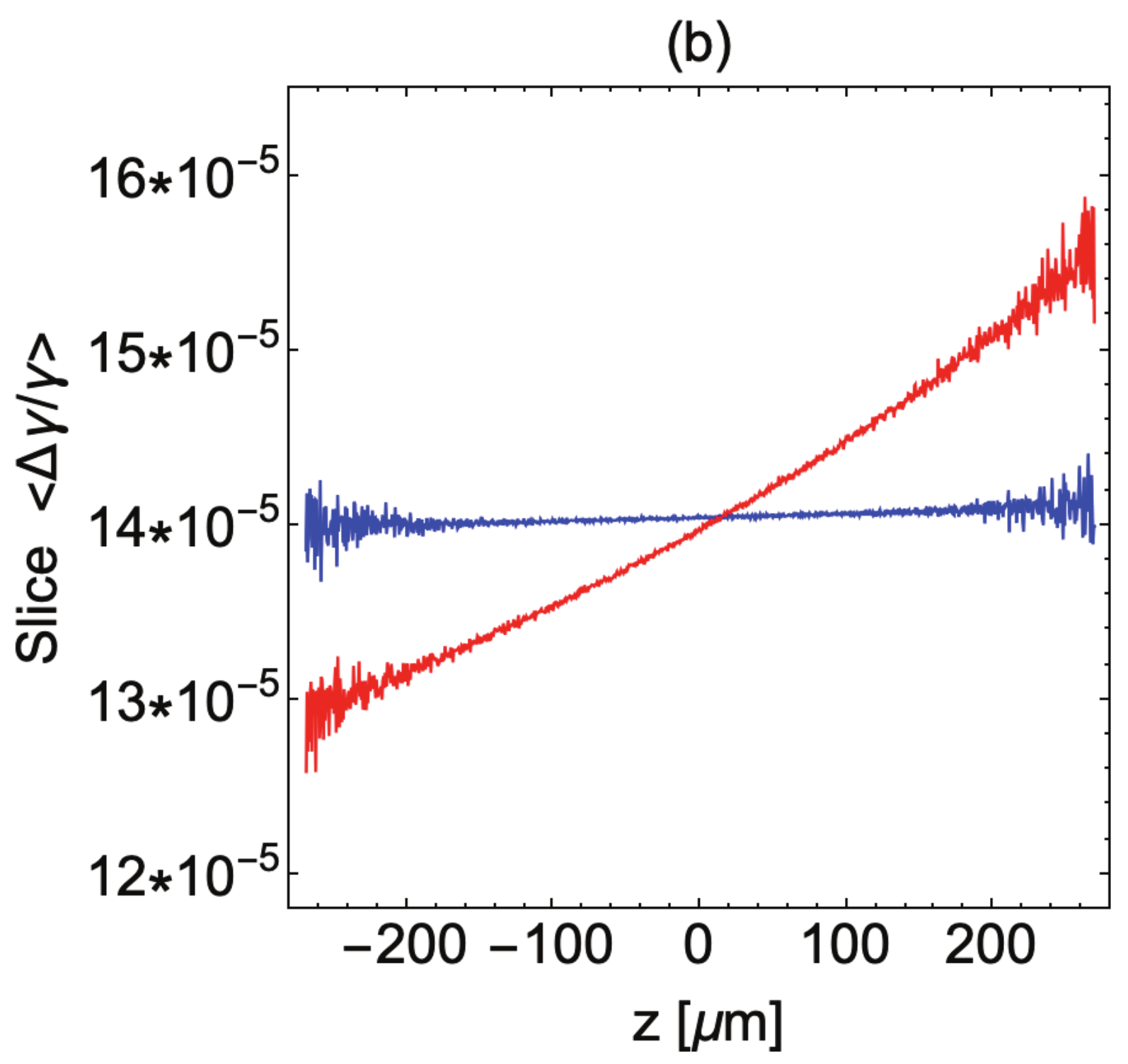}
	\caption{[Color online] Longitudinal phase space and slice energy spread at the exit of the TCBC at 250~MeV with tuned voltage $\Delta V$=+26.5~kV in the last TCAV and the optimized Twiss parameters of $\beta_{x,y}=150$~m and $\alpha_{x,y}=-1.3$ in the linear (blue) and nonlinear (red) regimes.}
	\label{fig:phase_space_250MeV_nosc_LastTDC_Twiss}
\end{figure}

The output longitudinal emittance $\epsilon_{nz}$ and corresponding eigen-emittance $\lambda_3$ are 31.86~$\mu$m, significantly larger than their initial values of 5.72~$\mu$m. The longitudinal phase space at the exit of the beamline is presented in Fig.~\ref{fig:phase_space_250MeV_nosc_LastTDC_Twiss}(a). The phase space shows the residual quadratic curvature similar to the one, observed during the off-crest acceleration. At the same time, the slice energy spread along the bunch is roughly the same both in
linear and nonlinear analysis of the TCBC beamline as illustrated in Fig.~\ref{fig:phase_space_250MeV_nosc_LastTDC_Twiss}(b). The curvature in the phase space can be compensated with the third harmonic cavity placed downstream in the beamline, similar to how it is done for conventional off-crest acceleration.

\begin{figure}[ht]
	\center
	\begin{tabular}{rr}
	&\includegraphics[height=2.0in]{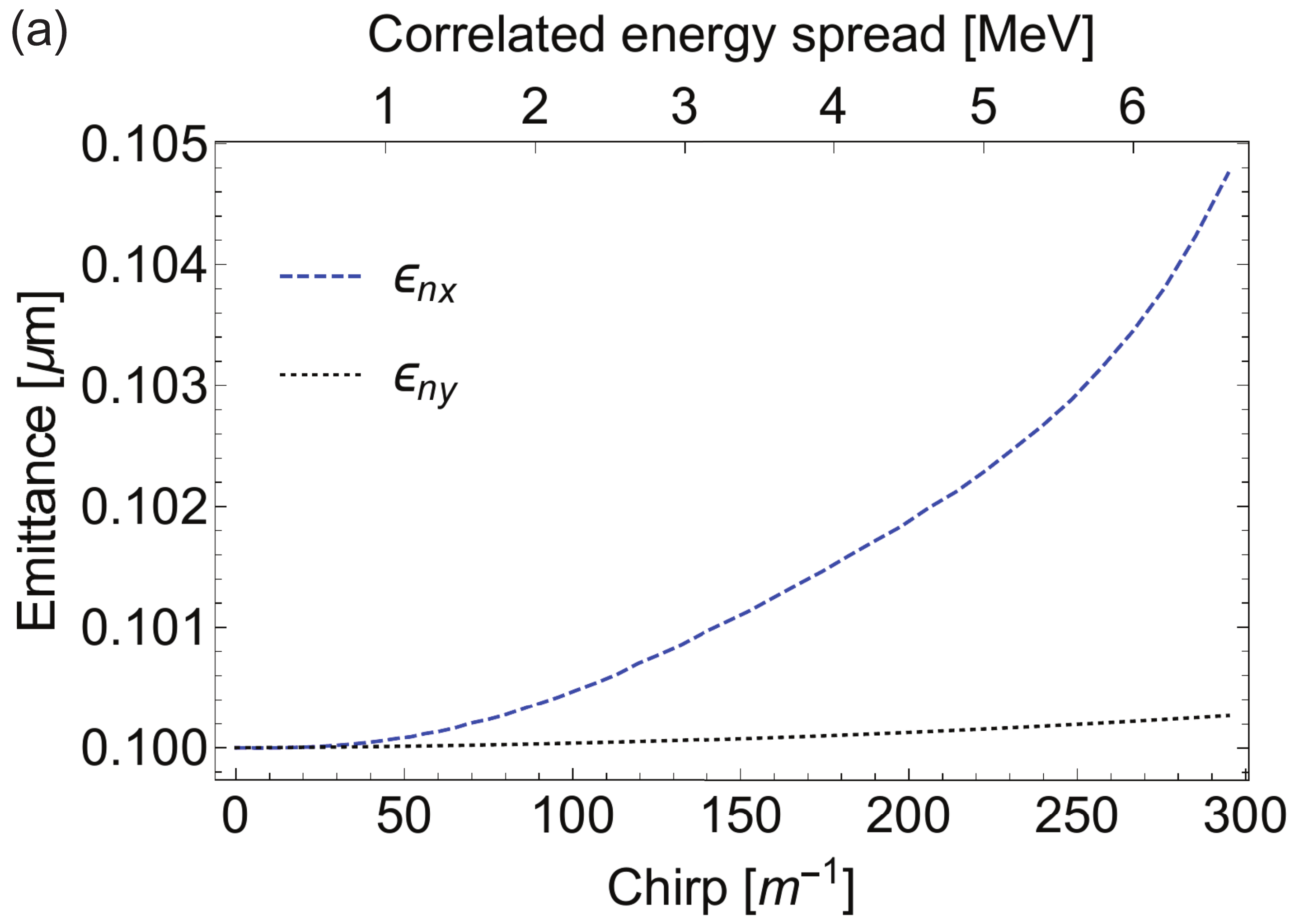}\\
	&\includegraphics[height=2.0in]{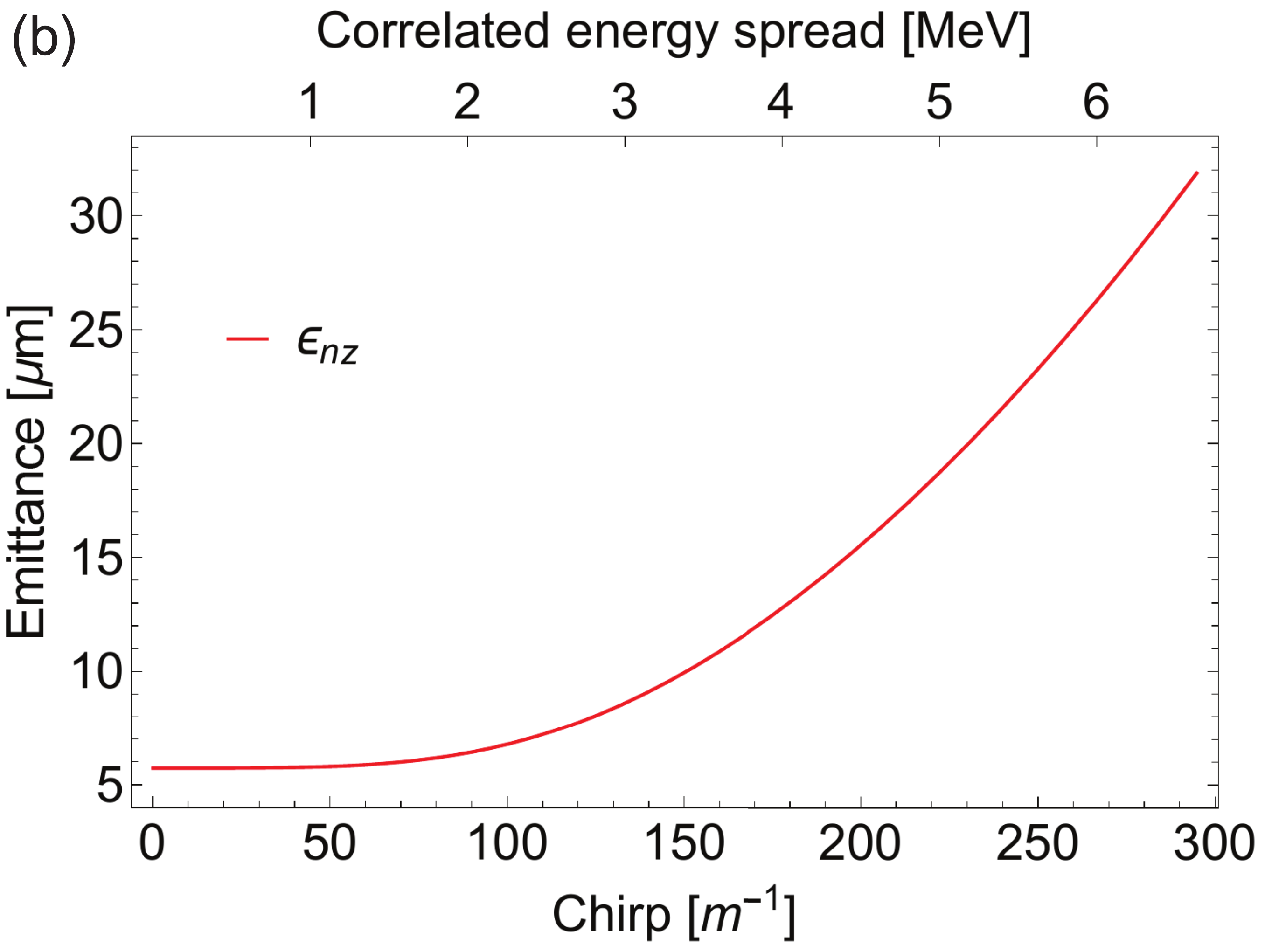}
	\end{tabular}
	\caption{[Color online] (a) Transverse  and (b) longitudinal emittances at the exit of the optimized TCBC at~250 MeV  vs the applied chirp.}
	\label{fig:Enx_Enz_250_Chirp}
\end{figure}

Figure~\ref{fig:Enx_Enz_250_Chirp} summarizes the performance of the optimized TCBC beamline. The emittance degradation depends on the applied chirp since it is caused by nonlinear effects which become more significant at larger values of imposed chirp. The imposed chirp is controlled by changing the deflecting voltage simultaneously in all cavities, while the voltage in the last cavity is adjusted according to the scaling in Eq.~(\ref{eq:scaling}). All other parameters of the scheme are not changed, including the geometry and the Twiss parameters (Twiss parameters have been optimized for the largest imposed chirp). Optimization of Twiss parameters for each deflecting voltage does not result in significant improvement in performance.

We emphasize that optimization of Twiss parameters and the voltage in the last cavity are equally important for the preservation of the beam quality. Optimization of Twiss parameters without proper adjustment to the TCAV4 voltage resulted in the smallest transverse emittance of $\epsilon_{nx}=0.2\;\mu{\rm m}$ as shown in Fig.\ref{fig:emittances_beta_a0}(c). On the other hand, tuning the voltage in TCAV4 without proper optimization in the Twiss parameters resulted in the smallest emittance of $\epsilon_{nx}=0.14\;\mu{\rm m}$ as shown in Fig.~\ref{fig:emittances_Last_TDC}(a). Simultaneous optimization of these parameters results in the transverse emittance of $\epsilon_{nx}=0.105\;\mu{\rm m}$ as shown in Fig.~\ref{fig:Enx_Enz_250_Chirp}(a).

\subsection{Chirper at 1~GeV.}
\label{sec:1GeV}

The design of the TCBC at 1~GeV fits 4 cryomodules and it is significantly longer than the design of the TCBC at 250~MeV which fits 2 cryomodules as illustrated in
Fig.~\ref{layout}. It also requires more RF power, due to larger length of the deflecting cavities assuming the same peak field
in TCAVs (See Table~\ref{tbl:chirper} for more details). This difference is driven by the unfavorable scaling of the TCAV strength with energy, $\kappa\propto 1/\gamma$, and
strong dependence of the imposed chirp on the cavity's strength, $R_{65}\propto\kappa^2$. As a result, the imposed correlated energy spread is smaller in high energy beams,
$\sigma_{\Delta \gamma}=\gamma m c^2 R_{65}\sigma_z\propto1/\gamma$. A longer beamline is required to impose the same correlated energy spread at higher beam energies.

\begin{figure}[ht]
	\center
		\begin{tabular}{rr}
	&\includegraphics[height=2in]{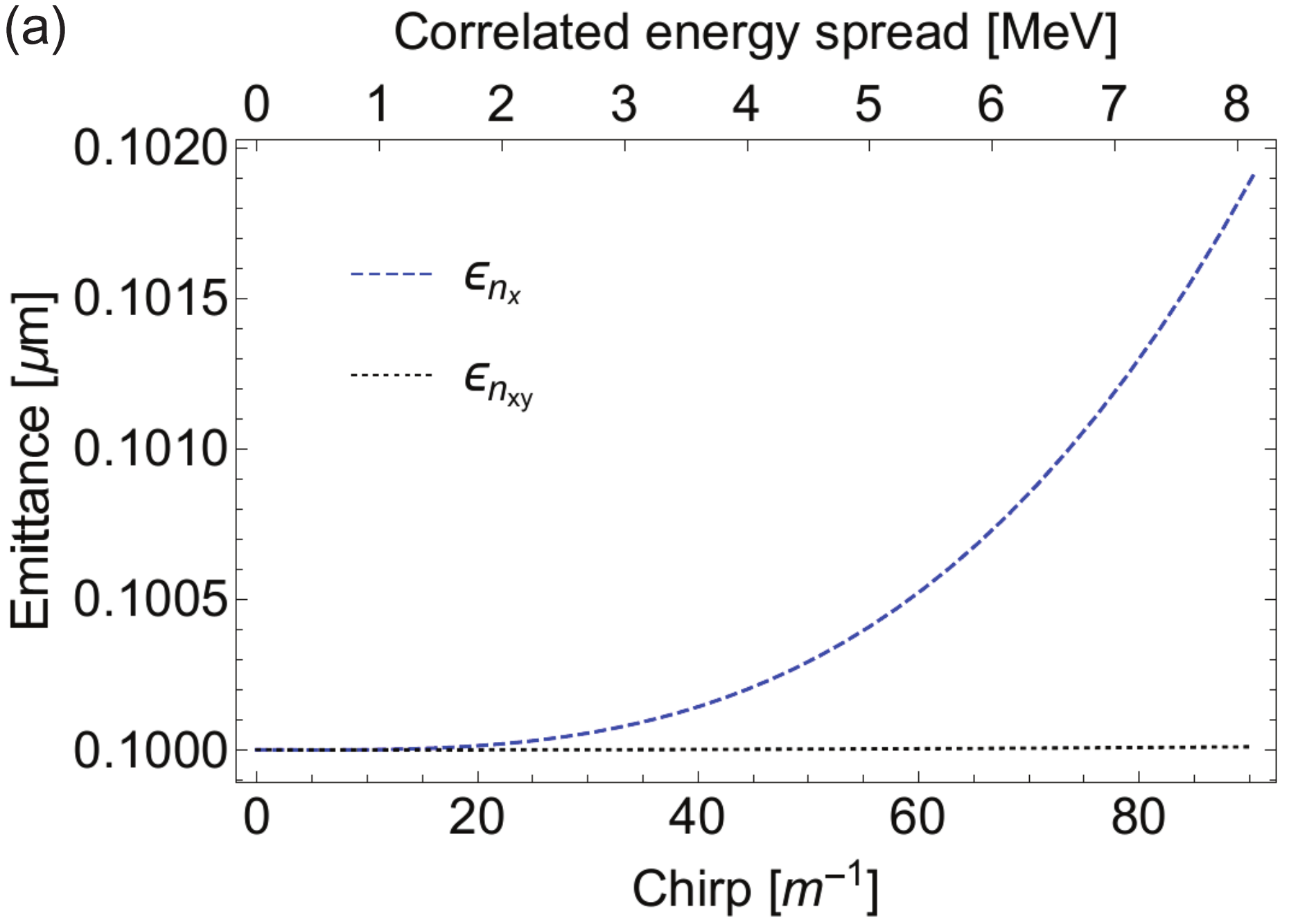}\\
	&\includegraphics[height=2in]{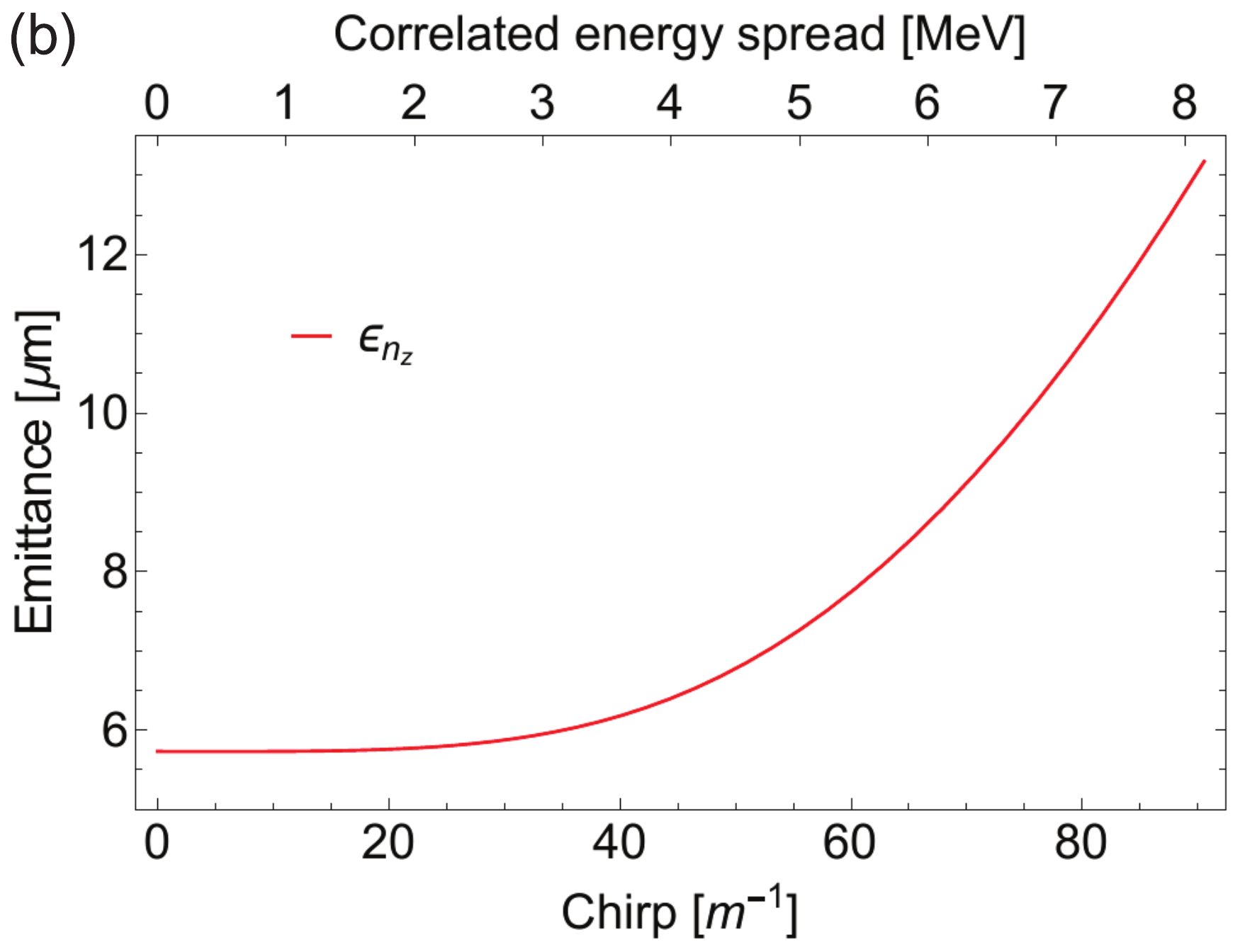}
		\end{tabular}
	\caption{[Color online] (a) Transverse  and (b) longitudinal emittances at the exit of the optimized TCBC at~1 GeV  vs the applied chirp.}
	\label{fig:Enx_Eny_1GeV_sc}
\end{figure}

At the same time, imposing the same correlated energy spread on the beam at higher energy results in smaller relative energy spread $\Delta \gamma/\gamma$, smaller nonlinear effects, and smaller residual transverse-to-longitudinal coupling. As a result, adjustment of the TCAV4 voltage is not required, in contrast to the TCBC at 250~MeV. The emittance growth is significantly smaller at higher energy as shown in Fig.~\ref{fig:Enx_Eny_1GeV_sc}. The optimized Twiss parameters in this case are
$\beta_x=175$~m and $\alpha_x=0.3$.

The designs of the TCBC at 250~MeV (Sec.~\ref{sec:250MeV}) and 1~GeV (\ref{sec:1GeV}) present a trade-off. The TCBC is more compact and cheaper at low energy. At the
same time, imposing chirp at higher energy results in the output beam of higher quality. Moreover, the space charge effects are reduced at higher beam energy. Imposing the chirp at larger energy is beneficial for suppressing the microbunching instability since the beam is accelerated faster. The choice of the energy for chirping the beam should be
determined from the start-to-end simulations of the linac thorough evaluation of the full degradation in the beam quality.

\section{TCAV-based dechirper}
\label{sec:Dechirper}

The TCAV-based chirper scheme studied in Sec.~\ref{sec:Linear_A} imposes chirp equal to $R_{65}=-\frac{2}{3}\kappa^2(3D+2L_c)$, which quadratically depends on the cavity's strength. As a result, the sign of the chirp cannot be switched by simply changing the voltage in TCAVs to the opposite sign. Formally, the $R_{65}$ matrix element can be positive if $|D|>\frac{2}{3}L_c$ and $D<0$. That can be achieved through the use of FODO lattice between the TCAVs, which beam transform matrix is identical to the drift space with negative drift length. Details on the design of the effective negative drift can be found in Appendix~\ref{App:negative_drift}

Inclusion of quadrupoles in the scheme preserves the concept of the transverse-to-longitudinal mixing, while the energy chirp is still predominantly changed in the middle cavity.
The proposed  {\bf T}ransverse deflecting {\bf C}avity {\bf B}ased {\bf D}echirper (TCBD) can be used to eliminate the residual chirp after compression in the final bunch compressor of the linac. The TCBD is an active element unlike the dechirper based on a wake excitation~\cite{Stupakov,KoreanExp,Antipov}. The performance of the scheme does not depend on the bunch charge and it is insensitive to the details of the longitudinal bunch profile. The amplitude of the imposed negative chirp is controlled with RF and can be quickly adjusted during operation.

\subsection{TCBD for MITS.}
\label{sec:DechirperMITS}

\begin{figure}[ht]
	\center
	\includegraphics[width=3.4in]{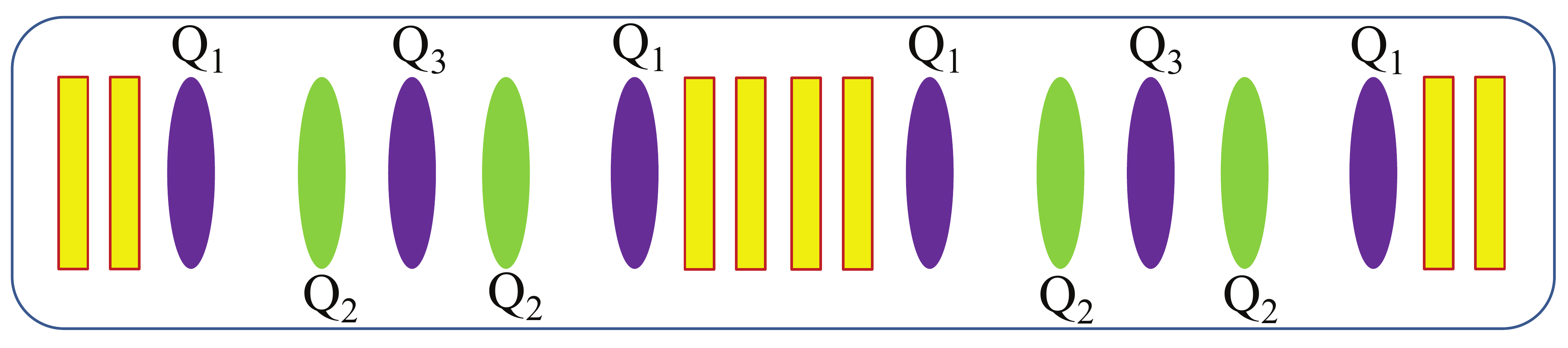}
	\caption{[Color online] The layout of TCAV-based dechirper inside the cryomodule for MITS. Each 13-cell cavity is represented in yellow, while the focusing and defocusing quadrupoles in respect to the ($x,\;x'$) phase space are in green and purple, respectively.}
	\label{fig:layout}
\end{figure}

Here we provide with a TCAV-based dechirper design for the MaRIE Injector Test Stand (MITS) for the beam energy of 250~MeV beam energy, which fits inside a single standard ILC cryomodule~\cite{ILC-cryo}. The superconducting quadrupoles with reasonable peak gradient are integrated in the scheme to provide the effective negative drift. The layout of the TCBD is shown in Fig.~\ref{fig:layout}. The main parameters are listed in Table~\ref{tbl:Dechirp_MITS}. The MITS dechirper at 250~MeV is designed to demonstrate the proof of principle of the concept, rather than being a practical application. Therefore, the proposed design will result in the 600~keV of the negative correlated energy spread which is sufficient for experimental demonstration.

\begin{table}[ht]
	\caption{Parameters of the TCAV-based dechirper beamline and beam for MITS}
	\vspace*{5mm}
	\centering{
		\begin{ruledtabular}
		\begin{tabular}{lll}
		 				&MITS  & MaRIE\\ 
			{\bf Electron beam }&   & \\ 
			Beam energy, MeV& 250  &1000\\
			Bunch charge, pC& 100 & 100\\
			Bunch length, $\mu$m & 90 & 3.9 \\
			Uncorrelated energy spread, keV& 32.5 & 750\\
			Normalized transverse emittance $\epsilon_{nx},~\epsilon_{ny}$, $\mu$m& 0.1  &0.1\\
			Normalized longitudinal emittance $\epsilon_{nz}$, $\mu$m& 5.72  &5.72\\
			Uncorrelated energy spread $\gamma mc^2\sigma_{\Delta\gamma/\gamma,u}$, keV&32.5& 750\\
			Correlated energy spread $\gamma mc^2 R_{65}\sigma_{z}$, MeV&0.6&5.2\\
			{\bf Beamline components}& & \\
			Total number of cryomodules & 1 & 6\\
			Total number of 13-cell cavities & 8 & 64\\
			Cavity frequency, GHz &3.9 & 3.9\\
			Effective RF length in each cavity, m &0.5  &0.5\\
			Deflecting voltage in each cavity, MeV&2.5 &2.5\\
			TCAV length $L_c$, m &1.38 &11.51\\
			TCAV strength $\kappa$, $\rm m^{-1}$ &0.82 & 0.205\\
			Total number of quads & 10 & 10\\
			Quadrupole length, m&0.05  &0.1\\
			Quadrupole 1  geometric strength, m$^{-2}$ & 23.8  &2.2\\
			Quadrupole 2  geometric strength, m$^{-2}$ & -28.4  &-5.0\\
			Quadrupole 3  geometric strength, m$^{-2}$ & 65.3 &16.5\\
			Effective negative drift, m & -5 & -65.3\\
			$R_{65}$, m & -26.7& -1333\\
			Beamline length, m&  13.1  & 75.9\\
			\end{tabular}
		\end{ruledtabular}}
	\label{tbl:Dechirp_MITS}
\end{table}

\begin{figure}[ht]
	\center
	\includegraphics[height=1.75in]{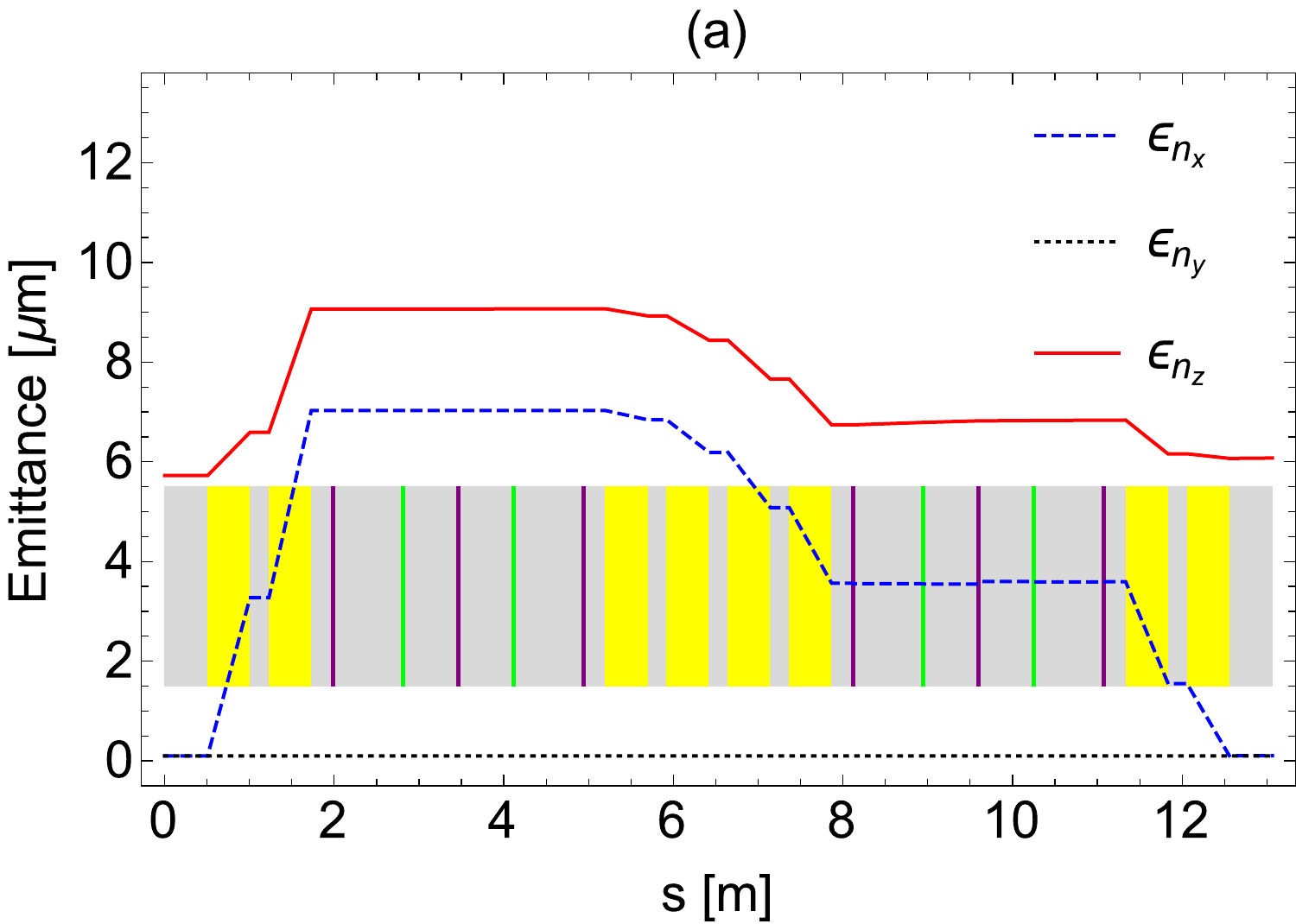}
	\includegraphics[height=1.75in]{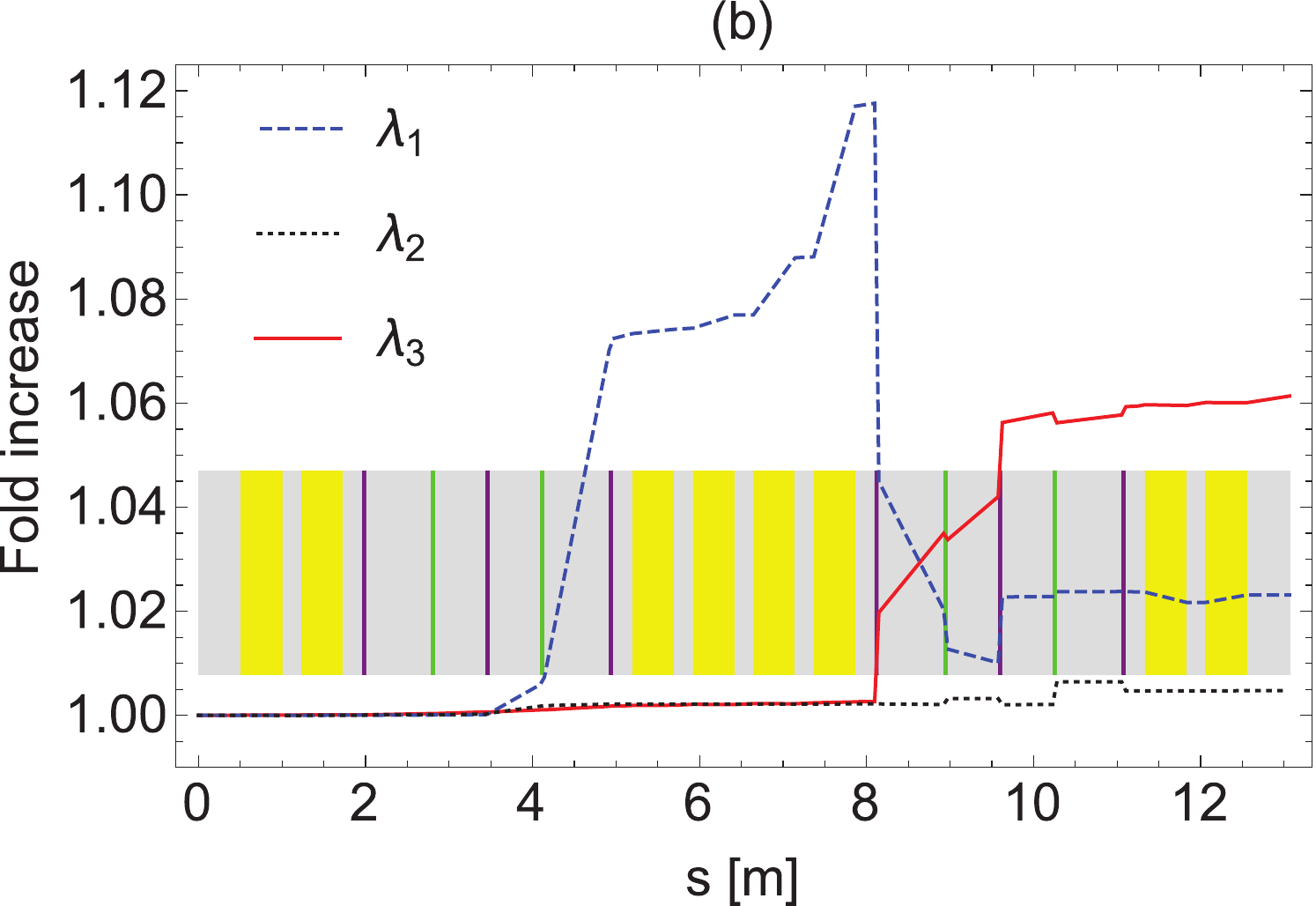}
	\includegraphics[height=1.75in]{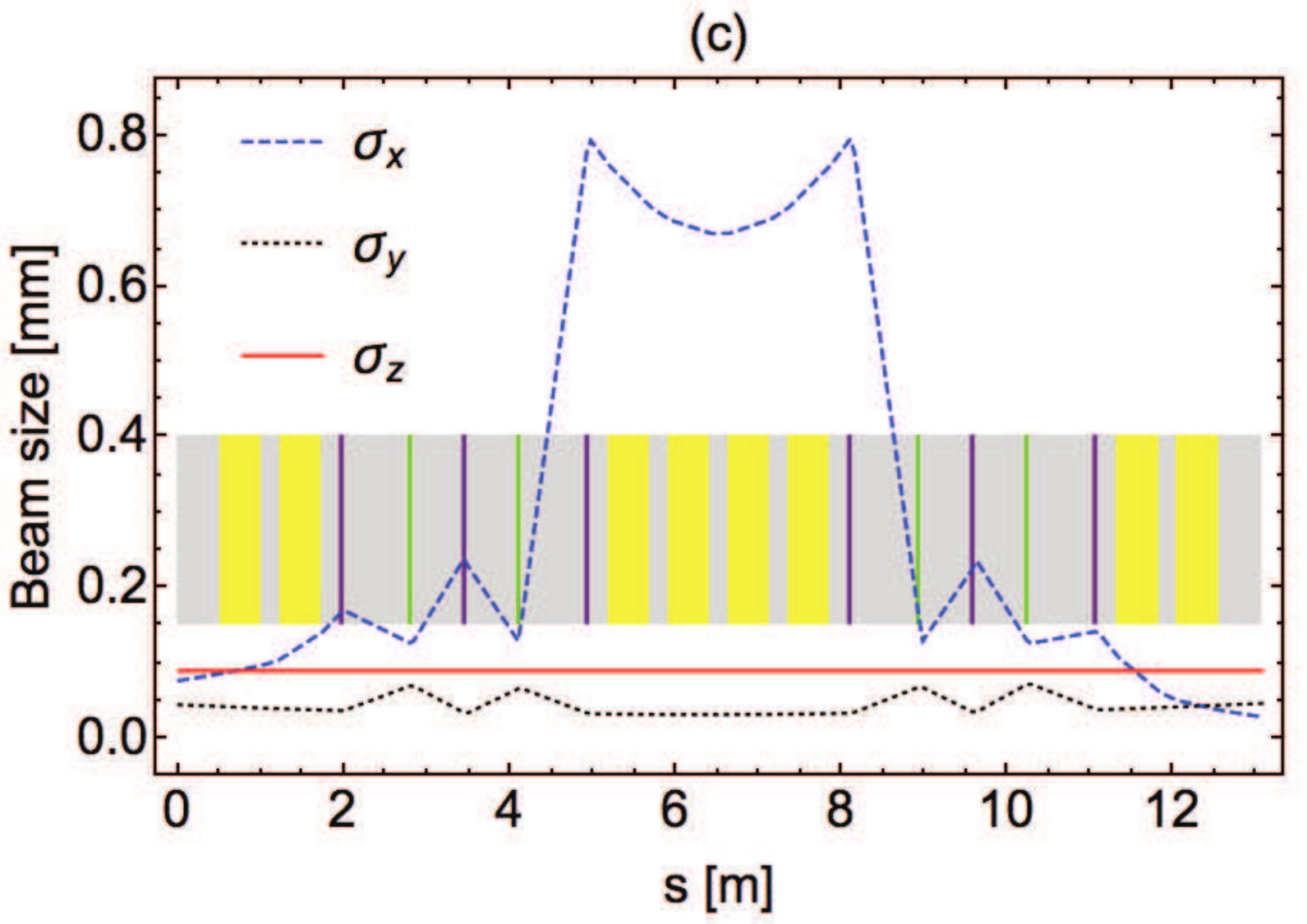}
	\includegraphics[height=1.75in]{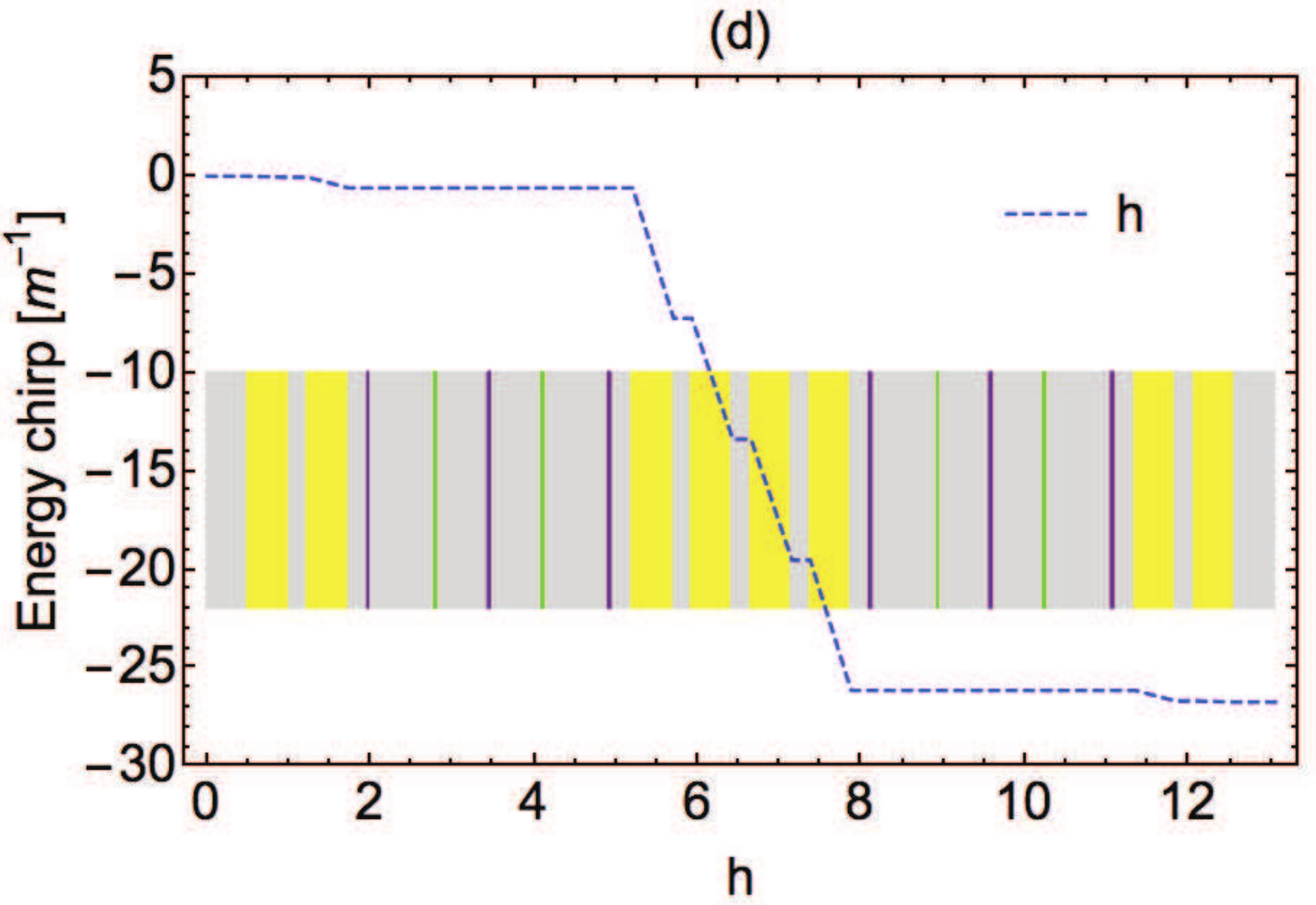}
	\caption{Evolution of the normalized emittances (a), related to them eigen-emittances (b), rms beam sizes (c), and energy chirp (d) at a distance $s$ along the beamline of the TCAV-based dechirper at 250~MeV for the Twiss parameters are $\beta_{x,y}=4.16$~m and $\alpha_{x,y}=-1$.}
	\label{fig:Emittances_MITS}
\end{figure}

The TCBD for MITS was analyzed in the approximation of the input beam with a zero chirp. Thus, the scheme provides with a decompressing energy chirp on the beam by the end of the beamline. The results are presented in Fig.~\ref{fig:Emittances_MITS}.
Growth of the transverse emittances can be reduced by adjusting the input Twiss parameters of the beam. 
In contrast to the chirper beamline, the adjustment of the Twiss parameters $\beta_x$ and $\alpha_x$ should be done independently from their counterparts corresponding to ($y,\;y'$) phase space for the dechirper beamline. The reason for that is the lack of symmetry in dynamics in x- and y- phase spaces. The optimization results in final transverse emittances of $\epsilon_{n_{x}}=0.1029\;\mu$m and $\epsilon_{n_{y}}=0.1003\;\mu$m, and a final longitudinal emittance of $\epsilon_{n_{z}}=6.08\;\mu$m for the optimal values of the input Twiss parameters of $\beta_x=28.8$~m $\beta_y=9.8$~m, $\alpha_x=-7$, and $\alpha_y=1$. The degradation of the beam quality due to nonlinear effects is not significant at the proper choice of Twiss parameters.

The proposed design results in the 600~keV of the decompressing energy chirp which is sufficient for experimental demonstration. At the same time, the imposed chirp is about a factor of 10 smaller
than in the chirper studied in Sec.~\ref{sec:250MeV}. The nonlinear emittance growth due to partial correlation between the longitudinal and the transverse phase spaces is not
significant for these parameters. We did not adjust the deflecting voltage in the last TCAV to minimize emittance growth and did not study emittance growth versus imposed chirp in this case.

\subsection{Dechirper design for the MaRIE linac at 1 GeV.}
\label{sec:Dechirper:MaRIE}

\begin{figure}[ht]
	\center
	\includegraphics[width=3.4in]{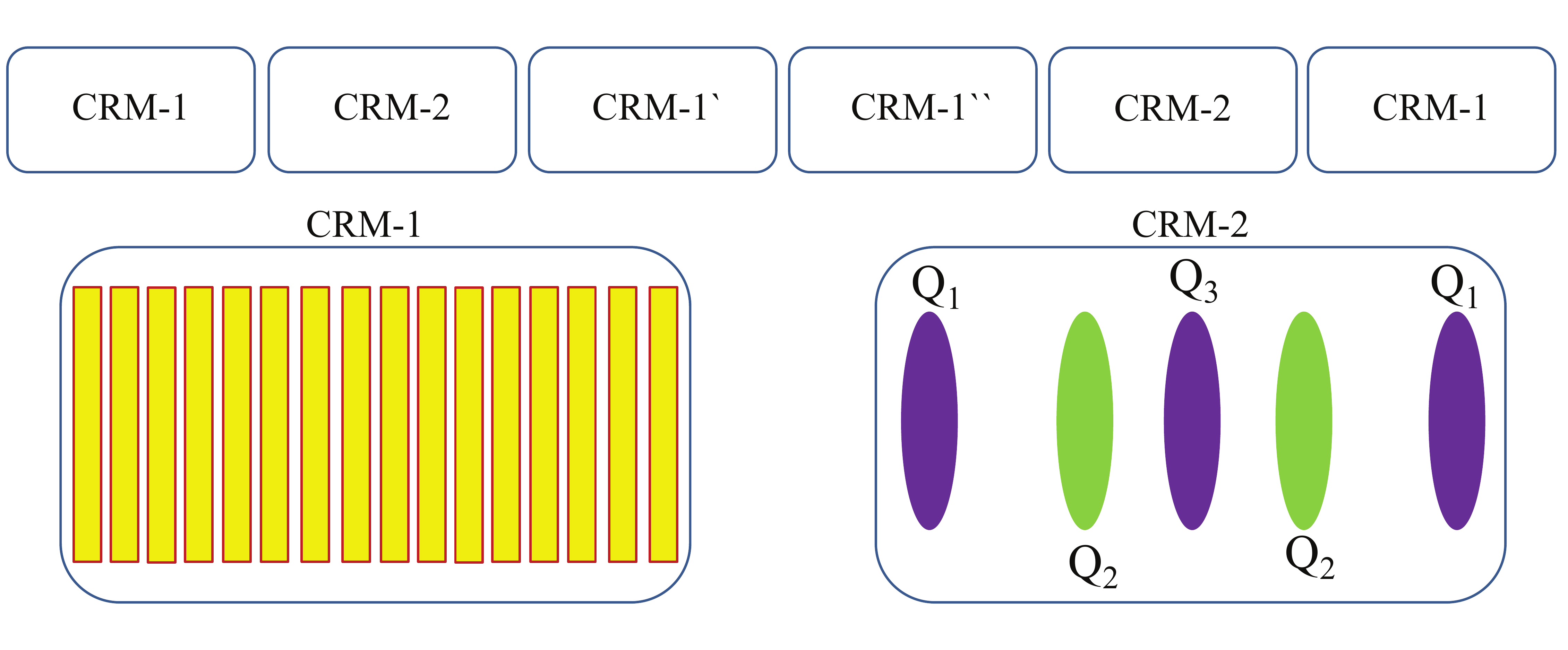}
	\caption{[Color online] The layout of TCAV-based dechirper at 1 GeV fitting inside six cryomodules. The cryomodules 1, 3, 4, and 6 have sixteen 13-cell cavities (shown in yellow). The cryomodules 2 and 5 host negative drift beamlines designed of focusing (green) and defocusing (purple) quadrupoles in respect to the ($x,\;x'$) phase space.}
	\label{fig:layout-1GeV}
\end{figure}

In Sec.~\ref{sec:DechirperMITS} we have presented a design for the TCBD for the future MITS facility. In this section we present a practical
design of the dechirper which may be implemented at MaRIE linac.
The overall design fits in six cryomodules as shown in Fig.~\ref{fig:layout-1GeV}. The length of the dechirper is significantly larger than than the chirper at 1 GeV
(Sec.~\ref{sec:1GeV}) since the bunch is compressed down to 3.9 $\mu$m compared to pre-compressed value of 90 $\mu$m (compression ratio of 23). The parameters of the beamline elements are listed in Table~\ref{tbl:Dechirp_MITS}. Identical negative drifts are implemented in the second and fifth cryomodules. Each negative drift consists of five quadrupole magnets separated by drifts. Note that the distances between quadrupoles (Fig.~\ref{fig:layout-1GeV}) and their strengths are different from those in a simple double triplet design described in Appendix~\ref{App:negative_drift}. The design of this negative drift represents a compromise between its efficiency and quality. It is compact enough to provide with the effective negative drift of 65~m without much degradation in the beam quality. A chirp (up to 5.2~MeV) was applied to the beam at the entrance to the scheme to simulate the beam coming out of the MaRIE chicane. The TCBD scheme was simulated as an actual dechirper and strengths of TCAVs were adjusted to eliminate the imposed chirp. 

\begin{figure}[ht]
	\center
	\includegraphics[height=1.75in]{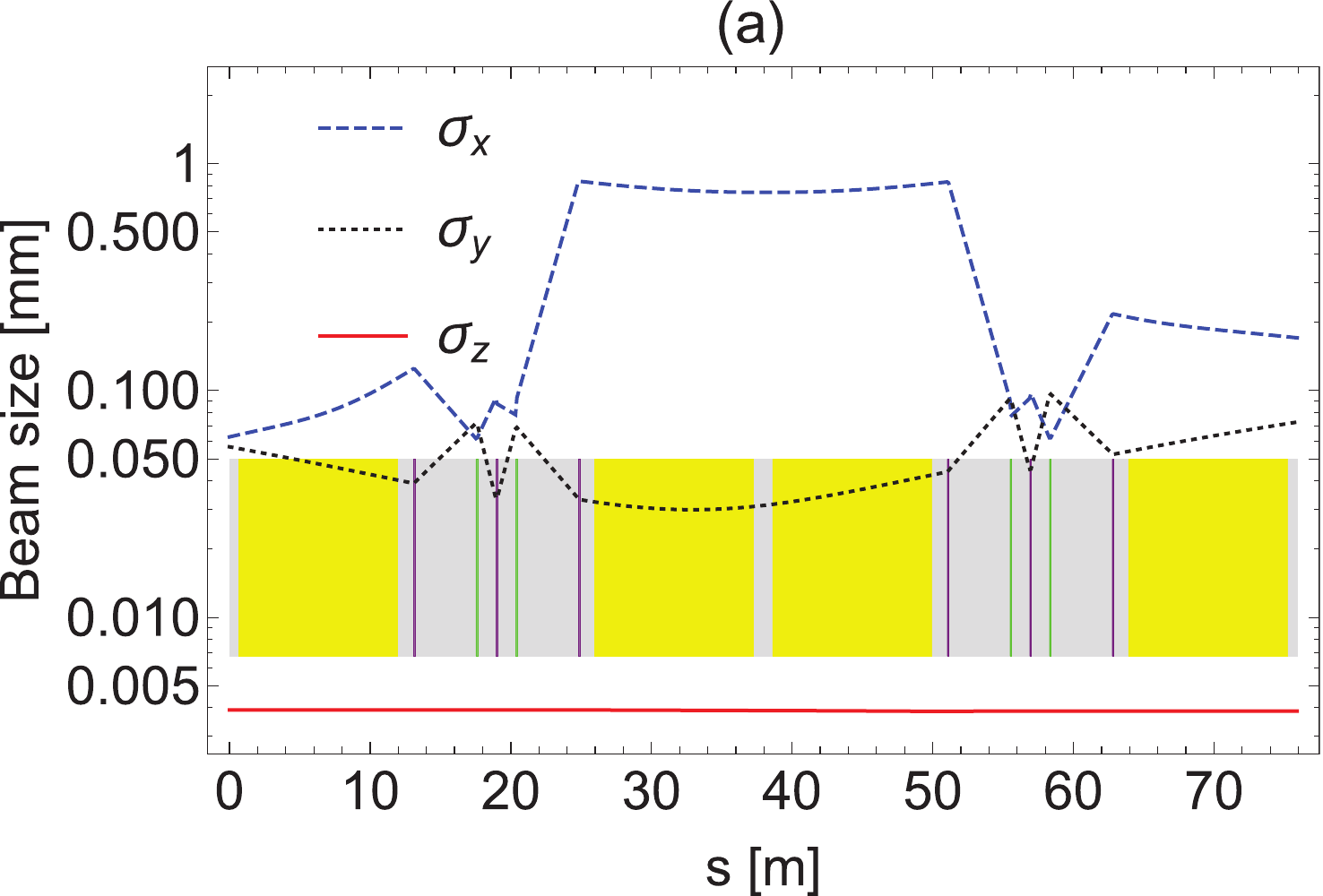}
	\includegraphics[height=1.75in]{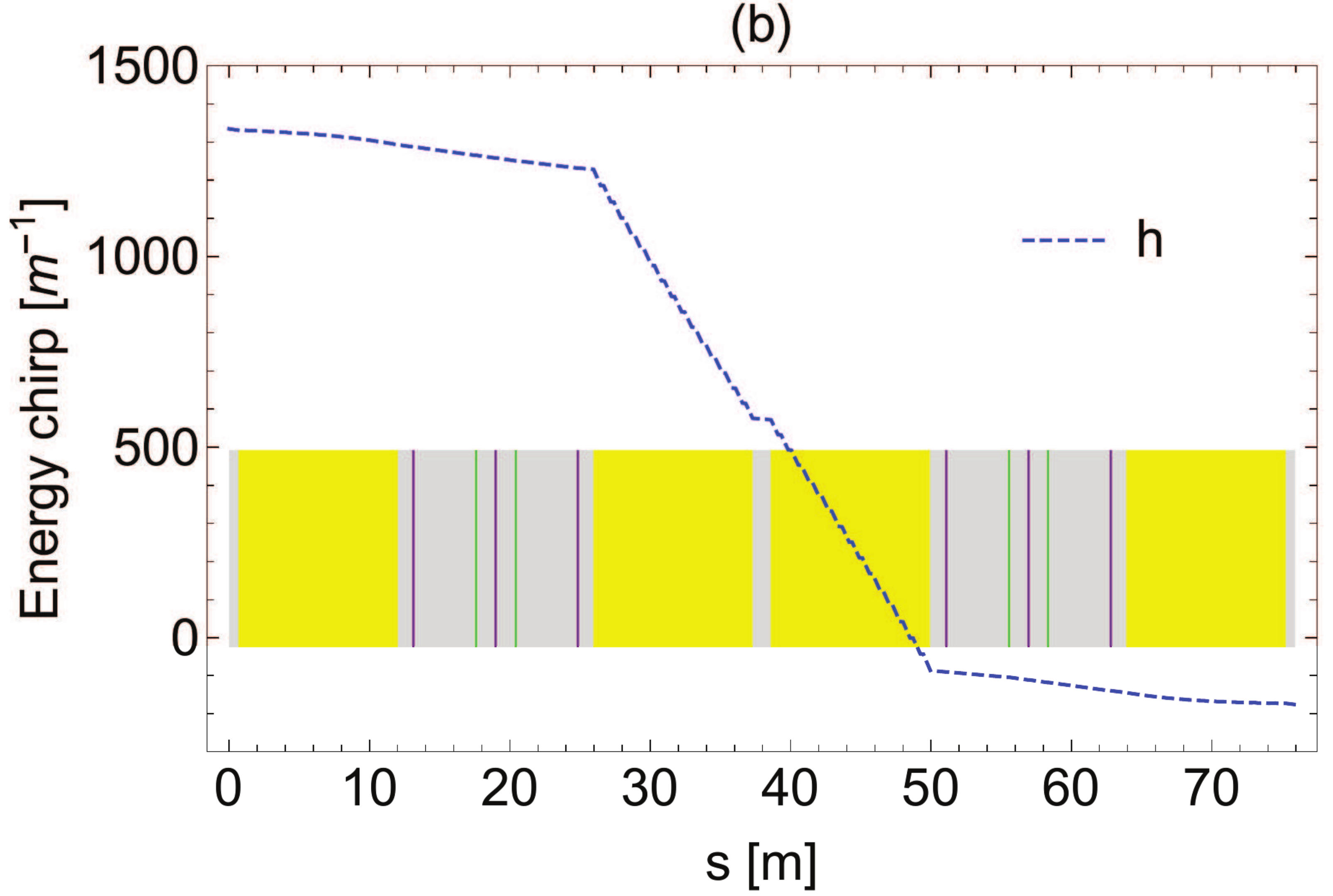}
	\includegraphics[height=1.75in]{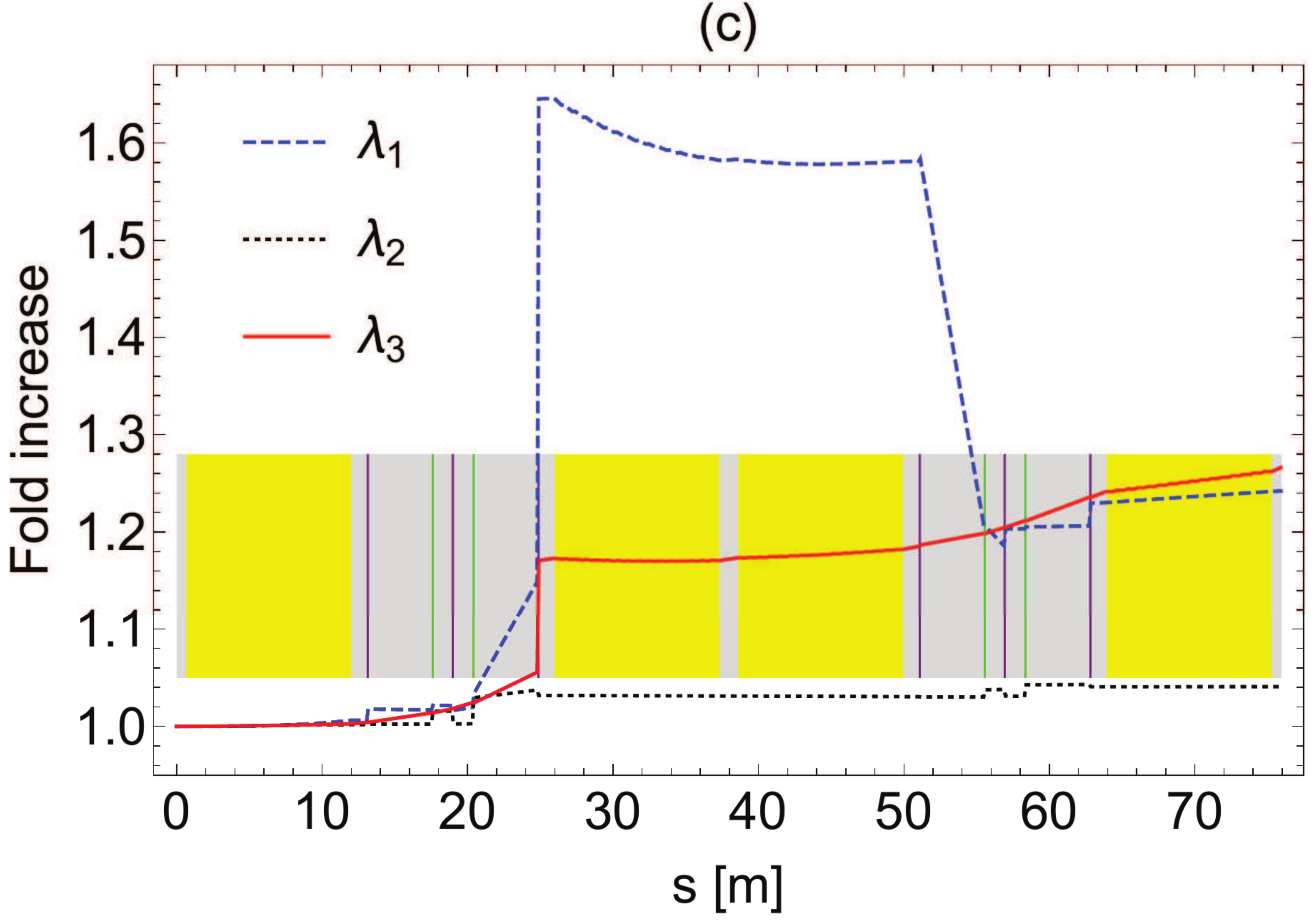}
	\caption{[Color online] Rms beam sizes (a); the total energy spread and chirp (b) and eigen-emittances (c), at a distance $s$ along the beamline of the TCAV-based dechirper at 1~GeV for Twiss parameters $\beta_x=76$~m, $\beta_y=62$~m, $\alpha_x=-2.5$, and $\alpha_y=1.6$.}
	\label{fig:Sx_Sy_dechirper-1GeV_Opt-Twiss}
\end{figure}

Figure~\ref{fig:Sx_Sy_dechirper-1GeV_Opt-Twiss} demonstrates the evolution of the beam parameters along the beamline. The chirp is mostly eliminated in the middle TCAV, similar to the TCBC scheme. The small residual transverse-to-longitudinal linear correlations are very small. If critical for larger normalized energy spread values, they can be adjusted by tuning the voltage in the last TCAV, similar to what was accomplished for the 250~MeV chirper scheme. We emphasize that this adjustment is opposite to the previous case: the voltage in the last cavity must be reduced compared to the linear case.
The Twiss parameters are optimized to minimize emittance growth. The output transverse emittances are $\epsilon_{nx}=0.127\;\mu$m, $\epsilon_{ny}=0.104\;\mu$m, and $\epsilon_{nz}=7.25\;\mu$m for the optimal Twiss parameters of $\beta_x=76$~m, $\beta_y=62$~m, $\alpha_x=-2.5$, and $\alpha_y=1.6$. 

\begin{figure}[ht]
	\center
			\includegraphics[height=2.0in]{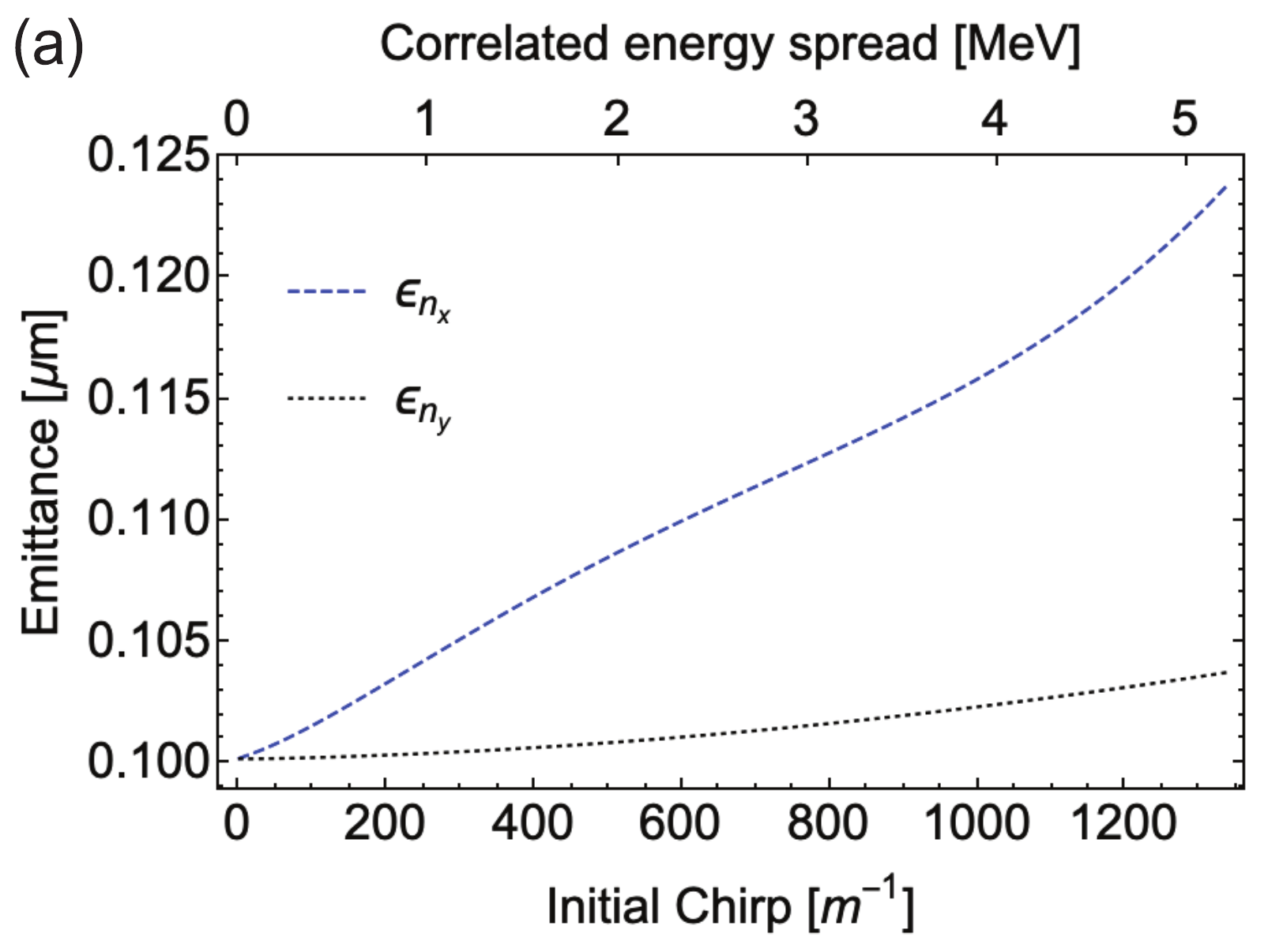}
			\includegraphics[height=2.0in]{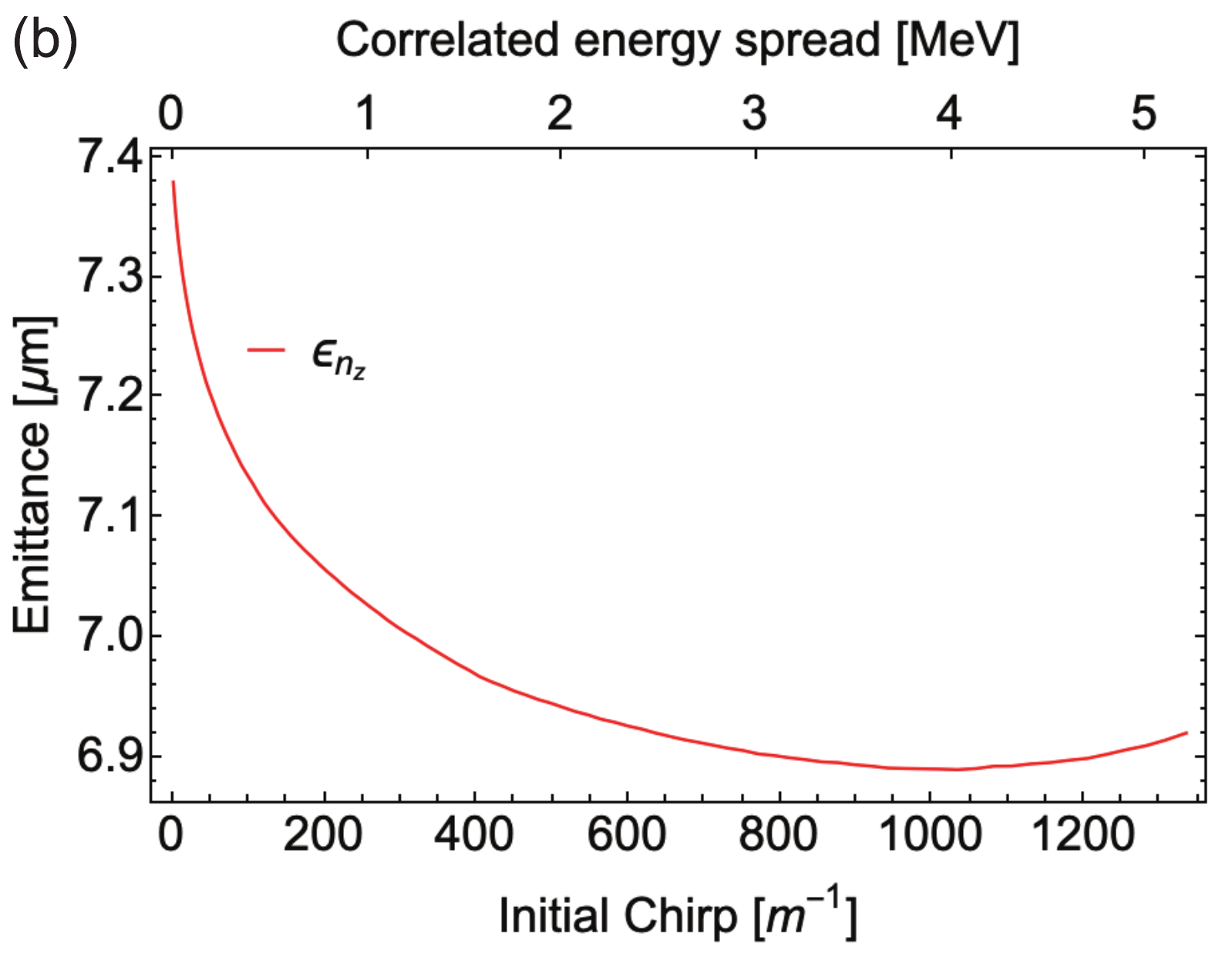}
			\includegraphics[height=2.0in]{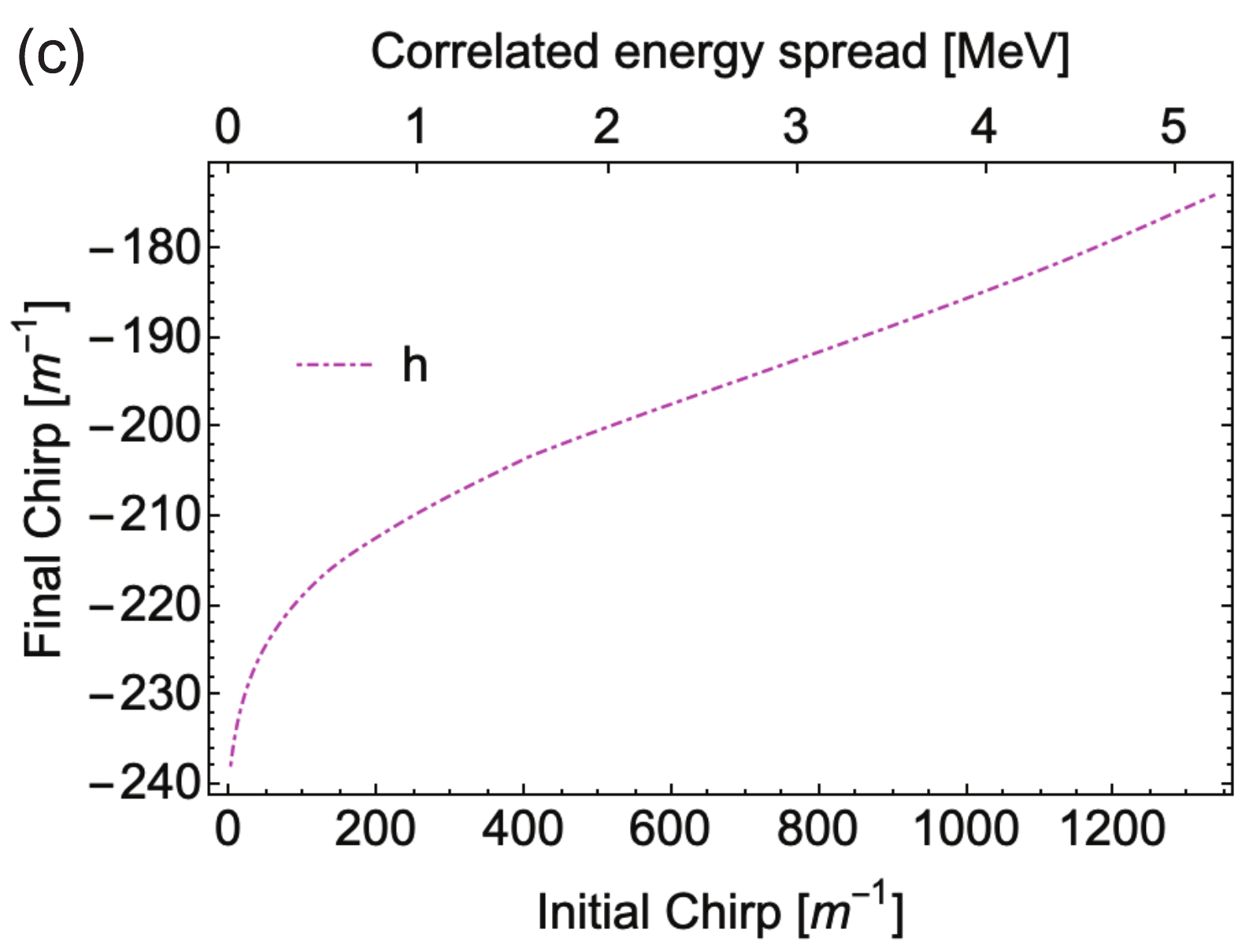}
		\caption{[Color online] (a) Transverse and (b) longitudinal emittances, and (c) chirp at the exit of the optimized TCBD at~1 GeV  vs the initial chirp for 100 pC bunch.}
	\label{fig:emittance-chirp-100pC}
\end{figure}

\begin{figure}[ht]
	\center
	\includegraphics[width=0.350\textwidth]{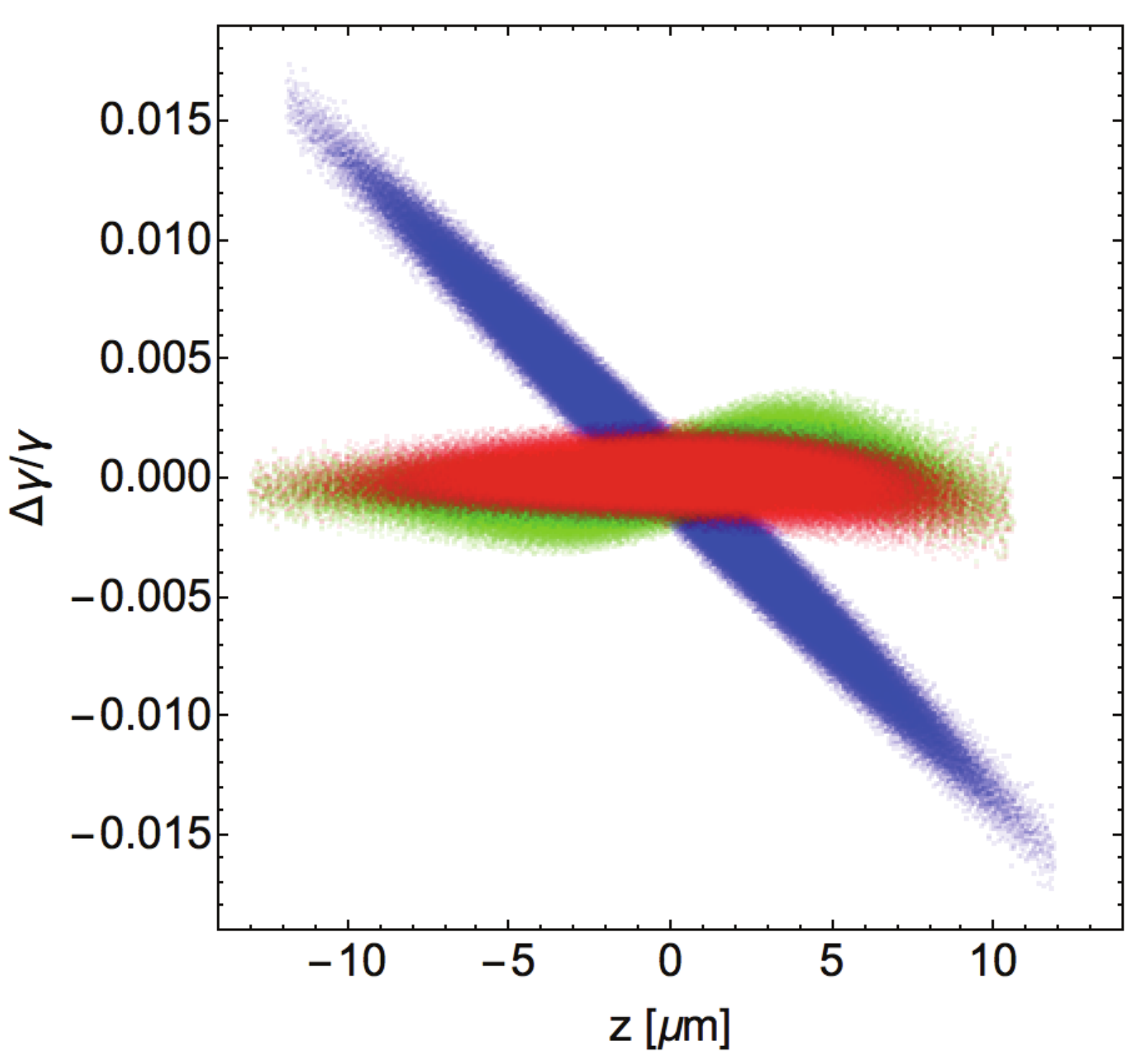}
	\caption{[Color online] Longitudinal phase space at the entrance (blue) and exit of the TCAV-based dechirper at 1 GeV with optimized input Twiss parameters for 0~pC (red) and 100~pC (green) bunches.}
	\label{fig:PS_Dechirp_1GeV_Opt-Twiss}
\end{figure}

The effect of the longitudinal space charge must be accounted for in the MaRIE TCBD design because of high peak current of $I=3$~kA after beam compression and relatively low beam energy. In fact, the beam propagates in a strongly space charge dominated regime after compression since $2I/I_A\frac{1}{(\beta\gamma)^3}\sigma_x^2\gg\epsilon_{nx}^2/\gamma^2$, where $\beta\approx1$ is the relativistic velocity (it is used only in this equation) and $I_A\approx17$~kA is the Alfven current. As a result, the emittances of the output beam are somewhat larger than what is expected for 0~pC bunch ($\epsilon_{nx}=0.127\;\mu$m, $\epsilon_{ny}=0.101\;\mu$m,  $\epsilon_{nz}=6.54 \;\mu$m). The transverse emittances are mostly not affected by the space charge but the longitudinal emittance grows due to the space charge wake. Figure~\ref{fig:PS_Dechirp_1GeV_Opt-Twiss} compares the longitudinal phase space at the entrance (blue) and exit of the beamline for 0~pC (red) and 100~pC (green) bunches.


Emittance growth and the final chirp are plotted versus initial chirp in 
Fig.~\ref{fig:emittance-chirp-100pC} for 100 pC bunch.

\section{Summary}
\label{sec:conclusion}

In this paper we have proposed and studied a novel scheme for imposing a linear energy chirp on an electron bunch using a combination of transverse deflecting cavities separated by drifts. 
We have designed two chirpers which can be placed at 250~MeV and 1~GeV, respectively. These schemes are designed to impose the required $>5$~MeV energy chirp prior to the bunch compressor
BC2. Both  chirpers are efficient and cause only minor degradation in the beam quality (see Fig.~\ref{fig:Enx_Enz_250_Chirp} and Fig.~\ref{fig:Enx_Eny_1GeV_sc}). Compensation of the higher-order nonlinear effects is required at the low energy of 250~MeV due to the large chirp compared to the energy of the beam. This was successfully accomplished through adjusting the voltage in the last transverse deflecting cavity and optimizing the input Twiss parameters. The design of the TCBC at a higher energy of 1~GeV results in a better quality beam but requires a longer length of RF structures and drifts. 

The TCBC concept can be modified for imposing decompressing energy chirp in the beam. Such a design can be used in conventional linacs for removing residual chirp in beams after the last bunch compressor. Such a design requires the use of FODO lattices effectively acting as negative drifts between deflecting cavities. Such an approach may be superior compared to passive dechirpers based on the excitation of longitudinal wakes inside of a corrugated pipe~\cite{Stupakov, Antipov, KoreanExp}. The main advantage of the proposed novel scheme
is a full control over the beam dynamics through beam optics elements. As a result, the TCBD does not depend on the bunch charge and the details of the current distribution (although, some degradation of the beam quality may be expected for high current bunches due to longitudinal space charge effects. We have presented the design of the TCBD at 250~MeV for the proof-of-principle demonstration at MITS and design for the TCBD dechirper for MaRIE linac at 1~GeV.

The TCBC and TCBD schemes are flexible and can be designed at different beam energies demonstrating a good performance. This suggests that they can be employed for essentially all existing and planned accelerators where it is required to impose or remove large energy chirps. This technique is superior to the conventional off-crest acceleration, particularly in terms of RF power consumption. The inexpensive option of imposing strong energy chirps may reduce the size of chicanes, which will mitigate degradation of the beam quality due to coherent synchrotron radiation (CSR).

\section{Acknowledgements}
One of the authors (AM) is grateful to Prof. Philippe Piot from Northern Illinois University and Dr. Paolo Craevich from Paul Scherrer Institute for fruitful discussions.

\appendix
\section{Negative drift}
\label{App:negative_drift}

Negative drift ($R_{12}\leq0$ and $R_{34}\geq0$) is a virtual element in beam optics and it can be designed as a sequence of focusing and defocusing quadrupole magnets separated by drifts. For example, an effective negative drift along the $x-$axis can be realized through as a combination of two triplets. Each triplet consists of two focusing ($f<0$) and one defocusing
($f>0$) quadrupoles which are separated by identical drifts ($d=-f$). The beam transport matrix describing a thin focusing/defocusing quadrupole acting in the 4D transverse phase space ($x,\;x',\;y,\;y'$):
\begin{equation}
M_{q_\pm}(f)=\begin{pmatrix}
1&0&0&0\\
\pm1/f&1&0&0\\
0&0&1&0\\
0&0&\mp1/f&1\\
\end{pmatrix}\;\;\;.
\end{equation}
The longitudinal beam dynamics is not affected by the quadrupole. 
The triplet matrix can be found as the product of its structural components:
\begin{equation}\label{eq:triplet_comb}
M_{t}(d)=M_{q_-}(d)\cdot M_{d}(d)\cdot M_{q_+}(d)\cdot M_{d}(d)\cdot M_{q_-}(d)\;\;\;,
\end{equation}
which results in:
\begin{equation}\label{eq:triplet_matrix}
M_{t}(d)=\begin{pmatrix}
-1&3d&0&0\\
0&-1&0&0\\
0&0&1&d\\
0&0&0&1\\
\end{pmatrix}.
\end{equation}
The combination of two triplets results in the effective negative drift for ($x,\;x'$) phase space and the positive drift for ($y,\;y'$) phase space: 
\begin{equation}\label{eq:neg_drift_as_triplet_comb}
M_{d^-}(d)=M_{t}(d)\cdot M_{t}(d)=\begin{pmatrix}
1&-6d&0&0\\
0&1&0&0\\
0&0&1&2d\\
0&0&0&1\\
\end{pmatrix}\;\;\;.
\end{equation}
The neighboring focusing elements of two triplets can be combined as a single quadrupole with a twice shorter focal length without loss of generality.



\begin{thebibliography}{8}
\bibitem{KEKB} 
S.~Kurokawa and E.~Kikutani, {\it Nucl. Instr. Meth. Phys. Res. A} {\bf 499} (1), 1, 2003.

\bibitem{CEBAF}
C.~W.~Leemann, D.~R.~Douglas, and G.~A.~ Krafft, Annu. Rev. Nucl. Part. Sci. {\bf 51}, 413 (2001).

\bibitem{ILC}
H.~Baer, T.~Barklow, K.~Fujii, Y.~Gao, A.~Hoang, S.~Kanemura, J.~List, H.~E.~Logan, A.~Nomerotski, M.~Perelstein {\it et al.}, ``The International Linear Collider Technical Design Report - Volume 2: Physics'', {\it arXiv:1306.6352}, (2013).

\bibitem{APS}
http://aps.anl.gov/

\bibitem{NSLS}
http://bnl.gov/


\bibitem{FLASH}
W.~ Ackermann, G.~Asova, V.~Ayvazyan, A.~Azima, N.~Baboi, J.~Bähr, V.~Balandin, B.~Beutner, A.~Brandt, A.~Bolzmann {\it et al.}, Nat. Photonics {\bf 1}, 336 (2007).

\bibitem{FERMI}
E.~Allaria, R.~Appio, L.~Badano, W.~Barletta, S.~Bassanese, S.~Biedron, A.~Borga, E.~Busetto, D.~Castronovo, P.~Cinquegran {\it et al.}, Nat. Photonics {\bf 6}, 699 (2012).

\bibitem{LCLS}
P.~Emma, R.~Akre, J.~Arthur, R.~Bionta, C.~Bostedt, J.~Bozek, A.~Brachmann, P.~Bucksbaum, R.~Coffee, F.-J.~Decker {\it et al.}, Nat. Photonics {\bf 4}, 641 (2010).

\bibitem{Spring-8}
T.~Ishikawa, H.~Aoyagi, T.~Asaka, Y.~Asano, N.~Azumi, T.~Bizen, H.~Ego, K.~Fukami, T.~Fukui, Y.~Furukawa {\it et al.}, Nat. Photonics {\bf 6}, 540 (2012).

\bibitem{XFEL}
T.~Tschentscher, C.~Bressler, J.~Grünert, A.~Madsen, A.~Mancuso, M.~Meyer, A.~Scherz, H.~Sinn, and U.~Zastrau, Appl. Sci. {\bf 7}, 592 (2017).

\bibitem{PAL-XFEL}
I~ Ko, H.-S.~Kang, H.~Heo, C.~Kim, G.~Kim, C.-K.~Min, H.~Yang, S.~Baek, H.-J.~Choi, G.~Mun {\it et al.}, Appl. Sci. {\bf 7}, 479 (2017).

\bibitem{SWISSFEL}
C.~J.~Milne, T.~Schietinger, M.~Aiba, A.~Alarcon  J.~Alex, A.~Anghel, V.~Arsov, C.~Beard, P.~Beaud, S.~Bettoni {\it et al.}, Appl. Sci {\bf 7}, 720 (2017).

\bibitem{CSR-JETP}
L.~V.~Iogansen, M.~S.~Rabinovich, Sov. Phys. JETP {\bf 37} (10), 83 (1960).

\bibitem{Bruce_Tor}
B.~E.~Carlsten and T.~O.~Raubenheimer, Phys. Rev.  E {\bf 51}, 1453 (1995).

\bibitem{Saldin_CSR}
 E.L.~Saldin, E.A.~Schneidmiller and M.V.~Yurkov, Nucl. Instrum. Methods Phys. Res., Sect. A
{\bf 398}, 373 (1997).

\bibitem{CSR}
R.~Li, Phys. Rev. ST Accel. Beams {\bf 11}, 024401 (2008).

\bibitem{LCLS_design}
J.~Arthur, P.~Anfinrud, P.~Audebert, K.~Bane, I.~Ben-Zvi, V.~Bharadwaj, R.~Bionta, P.~Bolton, M.~Borland, P.~H.~Bucksbaum {\it et al.},``Linac Coherent Light Source (LCLS) Conceptual Design Report'', SLAC National Laboratory Report SLAC-R-593, SLAC-R-0593, SLAC-593, SLAC-0593  (2002).

\bibitem{MARIE_update_2015}
J.~Lewellen, B.~Carlsten, F.~Krawczyk, Q.~Marksteiner, R.~Sheffield, and N.~Yampolsky , ``MaRIE X-FEL Linac Design: Status and Plans," {Los Alamos National Laboratory Report},
{ LA-UR-15-21962} (2015).


\bibitem{TCAV-diag}
A.~Scheinker, A.~Edelen, D.~Bohler, C.~Emma, and A.~Lutman,  {Phys. Rev. Lett.} {\bf 121}, 044801 (2018).



\bibitem{note1}
The restriction of having the same RF amplitude in all the cavities is redundant. It is sufficient that the side cavities have the same strength and the middle TCAV is twice stronger 
than the side ones. We assume the same ratios in their lengths for simplicity of the final expressions.

\bibitem{EEX}
M.~Cornacchia and P.~Emma, { Phys. Rev. ST Accel. Beams} {\bf 5}, 084001 (2002).

\bibitem{MaRIE}
http://marie.lanl.gov/

\bibitem{Saldin_Klystron}
E.~L.~Saldin, E.~A.~Schneidmiller, and M.~V.~Yurkov,  {Nucl. Instrum. Methods Phys. Res., Sect. A}
{\bf 490}, 1 (2002).

\bibitem{Huang_Microbunch_Inst}
Z.~Huang, M.~Borland, P.~Emma, J.~Wu, C.~Limborg, G.~Stupakov, and J.~Welch { Phys. Rev. ST Accel. Beams} {\bf 7}, 074401 (2004).

\bibitem{ILC-cryo}
T.J.~Peterson, M.~Geynisman, A.~Klebaner, V.~Parma, L.~Tavian, and J.~Theilacker, in ``{\it AIP Conference Proceedings} {\bf 985}, 1565 (2018). 


\bibitem{ILC-TCAV}
M.~McAshan and R.~Wanzenberg, Fermi National Accelerator Laboratory { FERMILAB-TM-2144} (2001).

\bibitem{Elegant}
M.~Borland, ``Elegant: A flexible sdds-compliant code for accelerator simulation,” { Advanced Photon Source Report LS-287} (2000).

\bibitem{noteLSC}
LSC effects' impact on the beam dynamics in the schemes located before BC2 becomes visible at 10~nC bunch charge.


\bibitem{Dragt}
A.~J~Dragt, J.~Neri, G.~Rangarajan, {\it Phys. Rev. A} {\bf 45} (4), 2572 (1992).

\bibitem{Dragt1}
A.~J.~Dragt, ``Lie methods for nonlinear dynamics with applications to accelerator physics," http://www.physics.umd.edu/dsat/dsatliemethods.html , (2001). 





\bibitem{Stupakov}
K.~L.~F.~Bane and G.~Stupakov, {Nucl. Instrum. Methods Phys. Res., Sect. A} {\bf 690}, 106 (2012).


 


\bibitem{KoreanExp}
P.~Emma, M.~Venturini, K.~L.~F.~Bane, G.~Stupakov, H.-S.~Kang, M.~S.~Chae, J.~Hong, C.-K.~Min, H.~Yang, T.~Ha {\it et al.}, { Phys. Rev. Lett.}  {\bf 112}, 034801 (2014).

\bibitem{Antipov}
S.~Antipov, C.~Jing, M.~Fedurin, W.~Gai, A.~Kanareykin, K.~Kusche, P.~Schoessow, V.~Yakimenko, and A.~Zholents, {Phys. Rev. Lett.}  {\bf 108}, 144801 (2012).





\end{thebibliography}
\end{document}